\RequirePackage{fix-cm}
\documentclass[twocolumn]{article}           % twocolumn

\usepackage{graphicx}
\usepackage{color}
\usepackage{amsmath}
\usepackage{subfigure}

\usepackage{amssymb}
\usepackage{natbib}

\usepackage[utf8]{inputenc}
\usepackage{textcomp}
\usepackage{overpic}

\usepackage{authblk}

\begin{document}

\title{Closed-loop separation control over a sharp edge ramp using Genetic Programming
}

\author[1]{Antoine Debien\thanks{antoine.debien@onera.fr}}
\author[2]{Kai~A.~F.~F. von Krbek\thanks{Kai.von.Krbek@krbek.de}}
\author[1]{Nicolas Mazellier}
\author[5]{Thomas Duriez}
\author[2]{Laurent Cordier}
\author[2,3]{Bernd R.\ Noack}
\author[4]{Markus W. Abel}
\author[1]{Azeddine Kourta}

\affil[1]{Universit\'e d'Orl\'eans, INSA-CVL, PRISME EA 4229, 8 rue L\'eonard de Vinci, F45072, Orl\'eans, France}%.......................................................................
\affil[2]{       
Institute PPRIME, CNRS -- Universit\'e de Poitiers -- ENSMA, UPR 3346,
D\'epartement Fluides, Thermique, Combustion,
CEAT, 43 rue de l'A\'erodrome,
F-86036 Poitiers Cedex, France}
%.......................................................................
\affil[3]{
Also at: Institut f\"ur Str\"omungsmechanik,
Technische Universit\"at Braunschweig,
Hermann-Blenk-Stra{\ss}e 37,
D-38108 Braunschweig, Germany}
%.......................................................................
\affil[4]{
Ambrosys GmbH, Albert-Einstein-Stra{\ss}e,
D-14473 Potsdam, Germany}
%.......................................................................
\affil[5]{
CONICET, 
Universidad de Buenos Aires, 
Ciudad Autonoma de Buenos Aires, Argentina.
Formerly at: Institute PPRIME, Poitiers, France
}

\maketitle
\begin{abstract}
We experimentally perform open and closed-loop control 
of a separating turbulent boundary layer  downstream from a sharp edge ramp.
The turbulent boundary layer just above the separation point 
has a Reynolds number $Re_{\theta}\approx 3\,500$
based on momentum thickness.
The goal of the control is to mitigate separation and early re-attachment.
The forcing employs a  spanwise array of active vortex generators. 
The flow state is monitored with skin-friction sensors downstream of the actuators.
The feedback control law is obtained using model-free genetic programming control (GPC)
(Gautier et al.\ 2015). 
The resulting flow is assessed using the
momentum coefficient, pressure distribution and skin friction over the ramp and stereo PIV.
The PIV yields vector field statistics, \textit{e.g.}\ shear layer growth, 
the back-flow area and  vortex region. 
GPC is benchmarked against the best periodic forcing.
While open-loop control achieves separation reduction by locking-on the shedding mode, 
GPC gives rise to similar benefits by accelerating the  shear layer growth.
Moreover, GPC uses less actuation energy.

%\keywords{Feedback Flow Control \and 
%                  Turbulent Boundary Layer \and 
%                  Active Vortex Generators \and 
%                  Machine Learning Control \and 
%                  Genetic Programming}


\end{abstract}

\section{Introduction}\label{sec:intro}
Fluid flows have an important impact on the performance of ground or airborne transport vehicles,
of gas and oil pipelines and of chemical and pharmaceutical processes, just to name a few applications.
Hence, the optimization of such flows by passive or active means to increase engineering performance
constitutes a very core discipline of fluid mechanics \citep{Brunton2015amr}.
%In  many aerodynamic applications, 
%the shape aerodynamic optimization has matured in the last 100 years of research. 
Passive flow control can provide an increase of performance without energy consumption.
For instance, vortex generators on the wings of most passenger aircrafts
prevent early separation and thus increase lift and reduce drag.
The effect of many passive devices can be emulated by active ones,
e.g.\ fluidic jets may act as active vortex generators.
Such active control can be turned on just in case of need,
or turned off to prevent parasitic drag. 
%or turned off to prevent parasitic drag, 
%or introduce frequency-dependent dynamics.
Active flow control is a key for further improvement for transport vehicles and combustion \citep{King2007book,King2010book}.
In particular, feedback control  can bring distinct benefits over blind open-loop forcing \citep{Rowley2006arfm},
e.g.\ can counteract instabilities in-time, perform disturbance rejection and compensate for model uncertainty.

The focus of this study is separation mitigation.
Separation occurs in many flow processes 
and may imply detrimental effects such as lift
reduction, drag increase and noise generation.
Downstream of the separation point, 
a shear layer develops and transports vorticity far from the wall. 
Large scale spanwise vortices emerge from the roll-up of the shear layer vorticity 
induced by the Kelvin-Helmholtz instability \citep{Mittal2005,Tennekes1972MIT}.
The Kelvin-Helmholtz instability is characterized by the Strouhal
 number $St_{\theta}=0.012$ based on the boundary layer momentum thickness 
$\theta$ close to the mean separation point \citep{Hasan1992JFM,Zaman1981JFM}. 
As the shear layer emerges  from the separation point, 
spanwise vortices progressively  grow and merge. 
This process leads to the shedding mode, 
characterized by a  Strouhal number $St_{L_{\textrm Sep}}=0.6 - 0.8$ 
\citep{Cherry1984JFM,Dandois2007JFM,Mabey1972JoA} based on the separation length 
$L_{\textrm Sep}$ and the free velocity flow field $U_{\infty}$.

Vortex generators (VGs) are a passive flow control device which 
may efficiently delay separation \citep{Lin2002PAS,Godard2006ASTP1}. 
VGs induce an array of streamwise vortices  energizing the near 
wall flow by momentum transfer with the free-stream. 
However, these VGs are permanently fixed to the system 
and come with a parasitic drag 
even in operating conditions where they are not needed.
Thus, these can induce drag penalties~\citep{Lin2002PAS} if 
flow control is just needed during specific phases. 

By contrast, active flow control, which adds energy in the flow using 
actuators, can be switched-off when control is not needed. Active 
separation control strategies are usually relying on optimal reduced frequencies 
$F^+$ to delay or shift a separation. In this paper, $F^+$ is scaled with
$L_{\textrm{Sep}}$ and $U_{\infty}$. According to literature, the optimal frequency 
range presents a large variability. For example, 
\citet{Seifert2003JoA} emphasized the efficiency of $0.5 \leq F^+ \leq 1.5$ 
while \citet{Greenblatt2000PAS} highlight an optimal 
reduced frequency range $2 \leq F^+ \leq 4$ and 
\cite{Amitay2002AIAA} show the possibility to use $F^+ \geq 10$. 

This large frequency range suggests to adapt the control
by closing the loop with sensors. 
Adaptive control approaches like extremum or slope seeking have shown good results for 
optimization \citep{Benard2010EiF,Shaqarin2013EiF}. 
However, an effective in-time control of large-scale coherent 
structures or dominant instabilities needs to respect 
the flow physics for the control design.
This design may be model-based using reduced-order models 
\citep{Gerhard2003aiaa,Pastoor2008JFM} to account for nonlinear actuation dynamics. 
Evidently, the control law depends on the quality of the flow model.
Such a model is still a challenge \citep{Cordier_Noack_Tissot_Lehnasch_Delville_Balajewicz_Daviller_Niven_2013_EIF}
for broad-band turbulence dynamics with frequency cross-talk
preventing a meaningful local linearization.

On the other hand, 
a model-free control design
using powerful methods of Machine Learning \citep{Murphy_2012}  
has been shown to be highly effective
in a number of experiments \citep{Duriez2014aiaa,Gautier2015jfm,Parezanovic2015ftac}.
This approach, called \textit{Genetic Programming Control} (GPC) in the sequel, 
detects and exploits nonlinear actuation mechanisms in an unsupervised manner.
GPC has outperformed the best open-loop control in terms of robustness 
and has worked even in case of a demonstrated nonlinear relation between actuators and sensors.
The key enabler is the application of genetic programming, a classical method of symbolic regression \citep{Koza1992book}, 
to optimize the closed-loop control law.
As such, GPC can be viewed as a generalization of the genetic algorithms often used to identify the parameters of control laws. 
We refer to the review article of  \cite{Brunton2015amr}
for an in-depth discussion of model-based and model-free turbulence control strategies.

In this paper, closed-loop control separation is performed using GPC 
and benchmarked  with an optimized open-loop control where vortex shedding is locked-on, 
leading to a reduction of the separation bubble~\citep{Debien2015TSFP9}.
The control is introduced by Active Vortex Generators (AVGs) which are  
set-up similarly
to the optimal configuration determined by \citet{Godard2006ASTP3} 
and \citet{Cuvier2011TSFP7b}.

The experimental setup including the description of the facility, the model, the 
actuators/sensors and measurement chains is detailed in Sec.~\ref{sec:exp_setup}. The GPC 
algorithm used for the control design is then described in Sec.~\ref{sec:MLC} with a 
special attention paid to the experimental implementation. In particular, we define the 
cost functional used to rank the different GPC individuals as a compromise between the 
reduction of the separation length and the value of the momentum coefficient needed to 
achieve it. In Sec.~\ref{sec:online_results}, results of the best open and closed-loop 
experiments are presented based on the measurements made during the GPC runs. At this 
stage, eight particular GPC individuals performing well during the learning process are 
highlighted and discussed.  To understand the mechanisms behind the best performing control 
laws, these individuals are further analysed in Sec.~\ref{sec:offline_eval} by including 
PIV measurements and by increasing the evaluation times used for characterizing the control 
laws. The performance of the different individuals is then evaluated in terms of the AVGs' 
characteristics, of the separation length and of the energetic impact on the flow.  
Finally, in Sec.~\ref{sec:controlled_flow_analysis}, the properties of the mixing layer 
resulting from one of the most efficient GPC individuals are benchmarked with the best 
open-loop in terms of vorticity 
thickness, turbulent kinetic energy, Reynolds stresses and distribution of vortex region 
area.

\section{Experimental Setup}
\label{sec:exp_setup}

\subsection{Wind tunnel and test section}
\begin{figure}
  \includegraphics[width=0.45\textwidth]{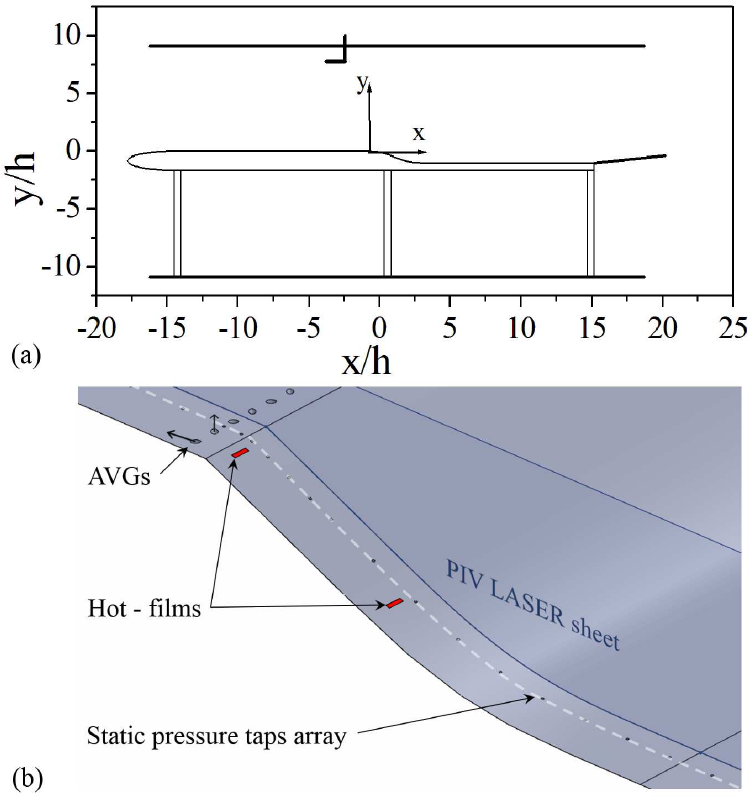}
\caption{a) Wind tunnel setup and b) close-up of the sharp edge ramp 
with the measurement facilities(not true to scale).}
\label{fig:RampSketch}
\end{figure}

The experiments are performed in the ``Lucien Malavard" subsonic wind tunnel
located at the PRISME Laboratory, University of Orl\'eans. The test
section is $2\,\mathrm{m}$ high, $2\,\mathrm{m}$ wide, and $5\,\mathrm{m}$ long.
The maximum free-stream velocity in the test section is $60\,\mathrm{m/s}$, and the residual turbulence intensity is below $0.4\%$. The ramp model (see Fig.~\ref{fig:RampSketch}) is set at the mid-height of the test section and spans the tunnel's width. 
The model is comprised of four parts: an elliptic leading edge, a flat plate that enables the development of a new, thin, boundary layer, a downward sloping ramp, and a second flat plate for the recovery region. Furthermore, a controllable flap is fixed at the trailing edge to control the stagnation point at the leading edge and to minimize the circulation around the model. This flap is set at an incidence of $7$\textdegree\ to ensure a symmetrical pressure distribution at the leading edge. The ramp has a length of $l = 470\,\mathrm{mm}$ and a step height of $h = 100\,\mathrm{mm}$.
The edge ramp is located at $x/h = 0$ with a slant angle of $25$\textdegree\ ending with a 7th order polynomial given by: 
\begin{equation}
\frac{y}{h} = 
1
-35\left(\frac{x}{l}\right)^4
+84\left(\frac{x}{l}\right)^5 
-70\left(\frac{x}{l}\right)^6 
+20\left(\frac{x}{l}\right)^7,
%\quad 0.5\leq x/l\leq 1
\label{eq:eq_trailing_edge}
\end{equation}
for $0.5\leq x/l\leq 1$.

For all the results presented in this paper, the free-stream velocity
is set to  $20\,\mathrm{m/s}$, 
achieving a Reynolds number $Re_\theta\approx 3\,500$ based on momentum thickness just above 
the sharp edge ramp. The boundary layer is tripped to fix the transition, thus warranting the 
reproducibility of its properties during the overall experiments. Further characterization of 
the un-actuated flow (hereafter called baseline) is provided in \citet{Debien2014CRAS}.

\subsection{Active vortex generators (AVG)}
\begin{figure}[htp]
  \includegraphics[width=0.45\textwidth]{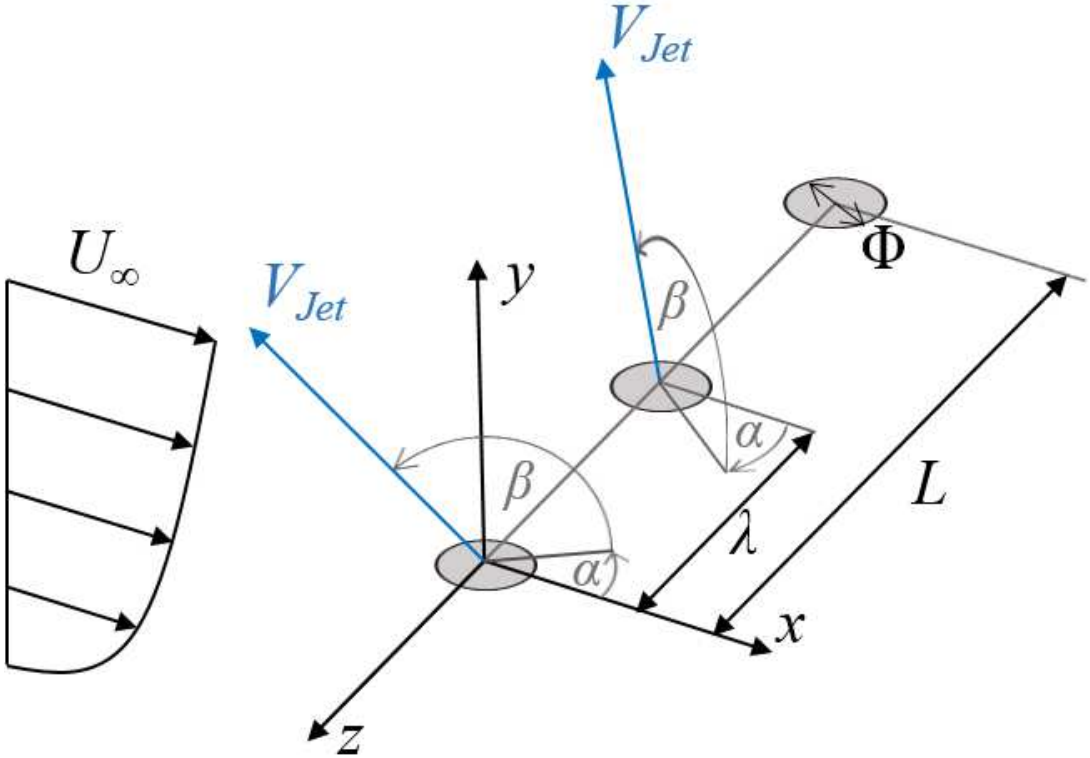}
\caption{Active vortex generators (counter-rotating configuration).
}
\label{fig:VGASketch}
\end{figure}

For introducing the control, $54$ AVGs in counter-rotating configuration are employed. Each AVG 
is composed of two jets (Fig.~\ref{fig:VGASketch}), 
leading to the generation of two counter-rotating  streamwise vortices that 
are triggered and driven in on/off mode.
These AVGs are positioned using the set of optimal parameters determined by
\citet{Godard2006ASTP3} and \citet{Cuvier2011TSFP7b}.
The AVGs are placed one boundary layer thickness upstream of the sharp edge ramp.
The diameter of the exit holes is $\Phi= 1.2\,\mathrm{mm}$.
The direction of the jets is characterized by the pitch and skew angles
$\alpha = 135$\textdegree\ and $\beta = 45$\textdegree, respectively. The distance between
two jets of the same actuator is $\lambda/\Phi = 15$, and the transverse distance
between the center line of two consecutive AVGs is $L/\Phi = 30$. The jets'
velocity ratio is regulated to be close to $V_{\textrm Jet}/U_{\infty} = 3$.

All the AVGs are supplied with air through a plenum chamber pressurized 
with compressed air coming from the Laboratory network. Due to the shape 
and size of the model, the plenum is divided into two sets of tanks. The 
first one of capacity $90\,\mathrm{L}$ is placed outside of the test 
section. In order to compensate for the pressure loss over the long tubing 
($4\,\mathrm{m}$) connecting the tank to the actuators, a set of three 
tanks of smaller capacity ($20\,\mathrm{L}$ each) is embedded in the 
profile, directly at the exit slots. Nine triggered electro-valves with 
sonic throats are connected to each of these three tanks. Each electro-valve 
supplies in on/off mode two pairwise AVGs, \textit{i.e.} a total of 4 jets.

The exit velocity of the jets is estimated using a calibrated $4^{th}$ 
order polynomial by measuring the pressure inside the second set of tanks. 
The pressure is measured by a differential pressure sensor 
($5\,\mathrm{bar}$, $\pm 100\,\mathrm{Pa}$, $0.1$\% FSO) and is regulated 
by a mass flow controller (Brooks, 5853 E series) providing up to 
$500\,\mathrm{L/min}$. A performance evaluation of this system showed that 
the pressure loss in the tank is less than $1.3$\% when AVGs are used 
in on/off mode with a constant actuation frequency $f_{\mathrm{Pulse}}$
in the range of $10\,\mathrm{Hz} \leq f_{\mathrm{Pulse}} \leq 
100\,\mathrm{Hz}$ (at 50\% duty cycle), corresponding to a velocity jet 
variation of 3\%.

Based on the determination of $V_\mathrm{Jet}$, the momentum coefficient 
$c_{\mu}$ is estimated as
\begin{equation}
c_{\mu} = 
\frac{\rho_{\mathrm{Jet}} 
S_\mathrm{Jet} 
D_c 
V_\mathrm{Jet}^2}
{\frac{1}{2}\rho_\infty S_{\mathrm{Ref}} U_\infty^2},
\label{eq:c_mu}
\end{equation}
where $\rho_\mathrm{Jet}$ is the flow density at the exit of the actuator, 
$\rho_{\infty}$ is the free-stream density, $D_c$ the duty cycle, 
$S_\mathrm{Jet}$ is the total cross section of the $108$ blowing jets, 
and $S_{\mathrm{Ref}}$ the streamwise projected area of the ramp
(height $\times$ spanwise length).

\subsection{Hot-film measurements}
For evaluating the performance of a given closed-loop control law, GPC needs to measure local flow characteristics. 
In the experiments, the wall shear stress at the ramp surface is measured on two locations (see Fig.~\ref{fig:RampSketch}) using hot-film probes.
These sensors give a monotonic signal of the absolute shear stress value \citep{Godard2006ASTP3,Cuvier2011TSFP7b,Shaqarin2011TSFP7},
which is directly 
induced by the local near wall velocity gradient. The two hot-films are
located within the recirculation bubble of the separation for the baseline flow.  
Due to the actuation, the separation length decreases leading to an
increase of the recirculation velocity within the recirculation
bubble. This could lead to a non-monotonic variation of the hot-film signals  
with respect to the separation length. In particular, if the separation length is decreased, a change in the 
hot-film signal leads to an ambigous interpretation. Indeed, the recirculation
velocity increases which leads to an increasing wall shear stress on the
sensors. If one is able to decrease the separation length even further,
the attachment point may be moved progressively towards the hot-film sensors
yielding a significant reduction of the wall shear stress. Thus, an evaluation of
the control effectiveness based solely on the hot-film sensor value is not
sufficient. Therefore, a set of static pressure sensors located along the span are also used to assess the effectiveness of the
control (see Fig.~\ref{fig:RampSketch}).

Senflex SF9902, made of active Nickel elements, are used as hot-film probes.
These sensors are $1.5\,\mathrm{mm}$ long, $0.102\,\mathrm{mm}$ wide, and are deposited on a polyamyde substrate with a
thickness less than $0.2\,\mathrm{mm}$. 
They are glued directly on the model's surface with $76\,\mu\mathrm{m}$ double-sided tape at 
$x/h = 0.06$ and $x/h = 1.38$ (Fig.~\ref{fig:RampSketch}).
The conditioning of the signal is achieved using a Dantec 90H02 Flow Unit.
The signal from the anemometer was low-pass filtered at $300\,Hz$
and conditioned on a $0-3.3\,\mathrm{V}$ range before feeding the signal into an Arduino. In section \ref{sec:MLC}, this Arduino will be tasked to calculate in real-time the actuation based on the data of the hot-film sensors.

\subsection{Pressure measurements}
The pressure distribution along the ramp model is obtained using a PSI 8400  
acquisition unit ($2500\,\mathrm{Pa}$, $\pm 0.75\,\mathrm{Pa}$) which allows the measurement of the $80$ 
pressure transducers inserted into the model.
The pressure taps ($0.3\,\mathrm{mm}$ in diameter) are connected to pressure sensors by a $1.5\,\mathrm{m}$ long tygon 
capillary.
Time-series of fluctuations pressure are acquired with a $200\,\mathrm{Hz}$ sampling frequency. 
During the GPC process, the pressure distribution over the ramp 
($0 \leq x/h \leq 4.5$) is measured using a recording time of $1\,000\,U_{\infty}/h$, corresponding to an uncertainty in the estimation of the 
pressure coefficient
($C_p = \frac {P-P_{\infty}}{1/2 \rho_{\infty} U_{\infty}^2}$) of $\pm 7.5\%$.
The pressure distribution for the baseline, the open-loop case and the 
best GPC individuals is then obtained using a recording time of $90\,000\,U_{\infty}/h$. This allows to determine the pressure coefficient with an uncertainty 
estimation of $\pm 1\%$.

\begin{figure*}[tp]
\centering
\subfigure[\label{fig:VtxInit}At the beginning.]{
\includegraphics[width=0.45\textwidth]{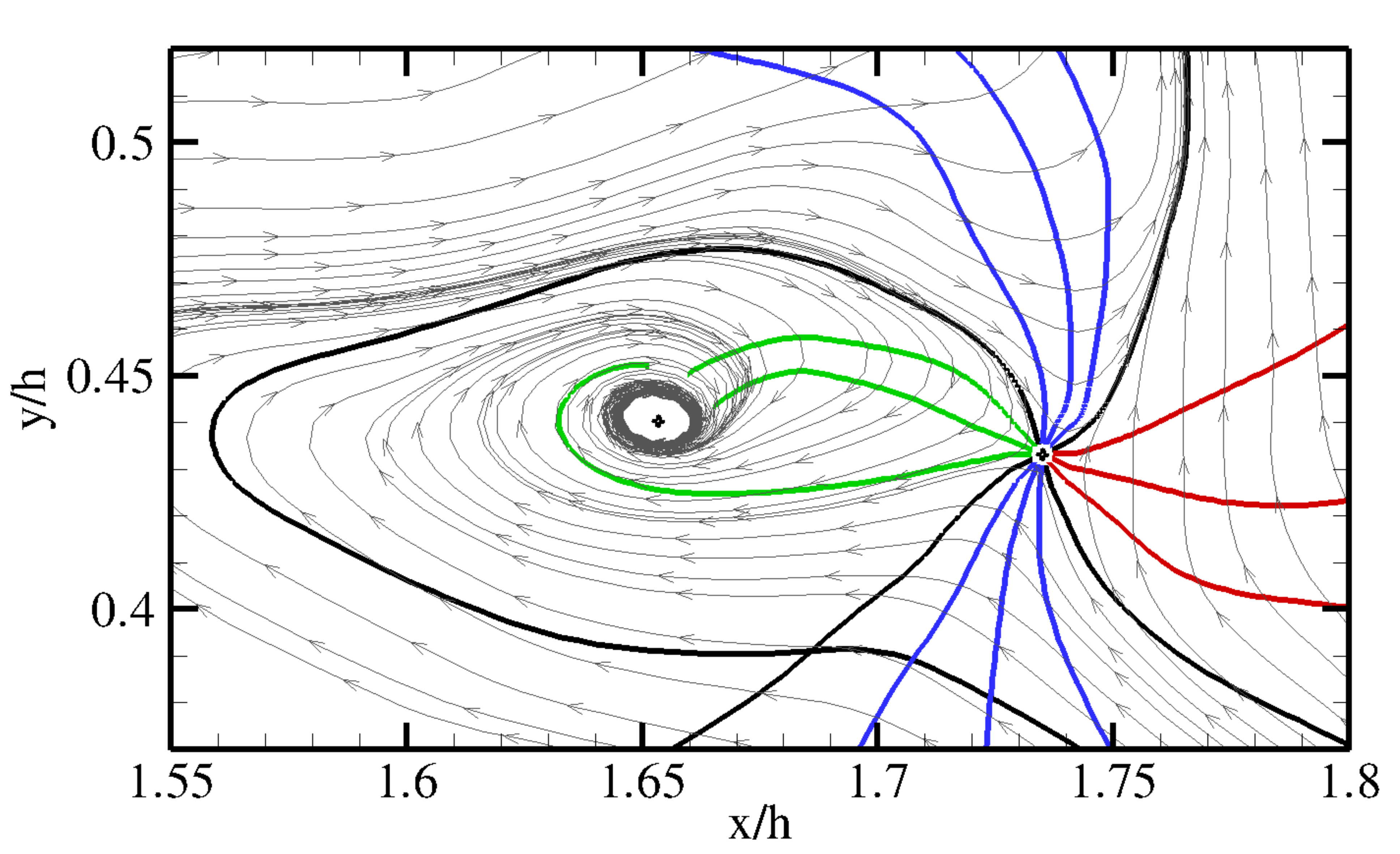}
}
\subfigure[\label{fig:VtxEnd}At convergence.]{
\includegraphics[width=0.45\textwidth]{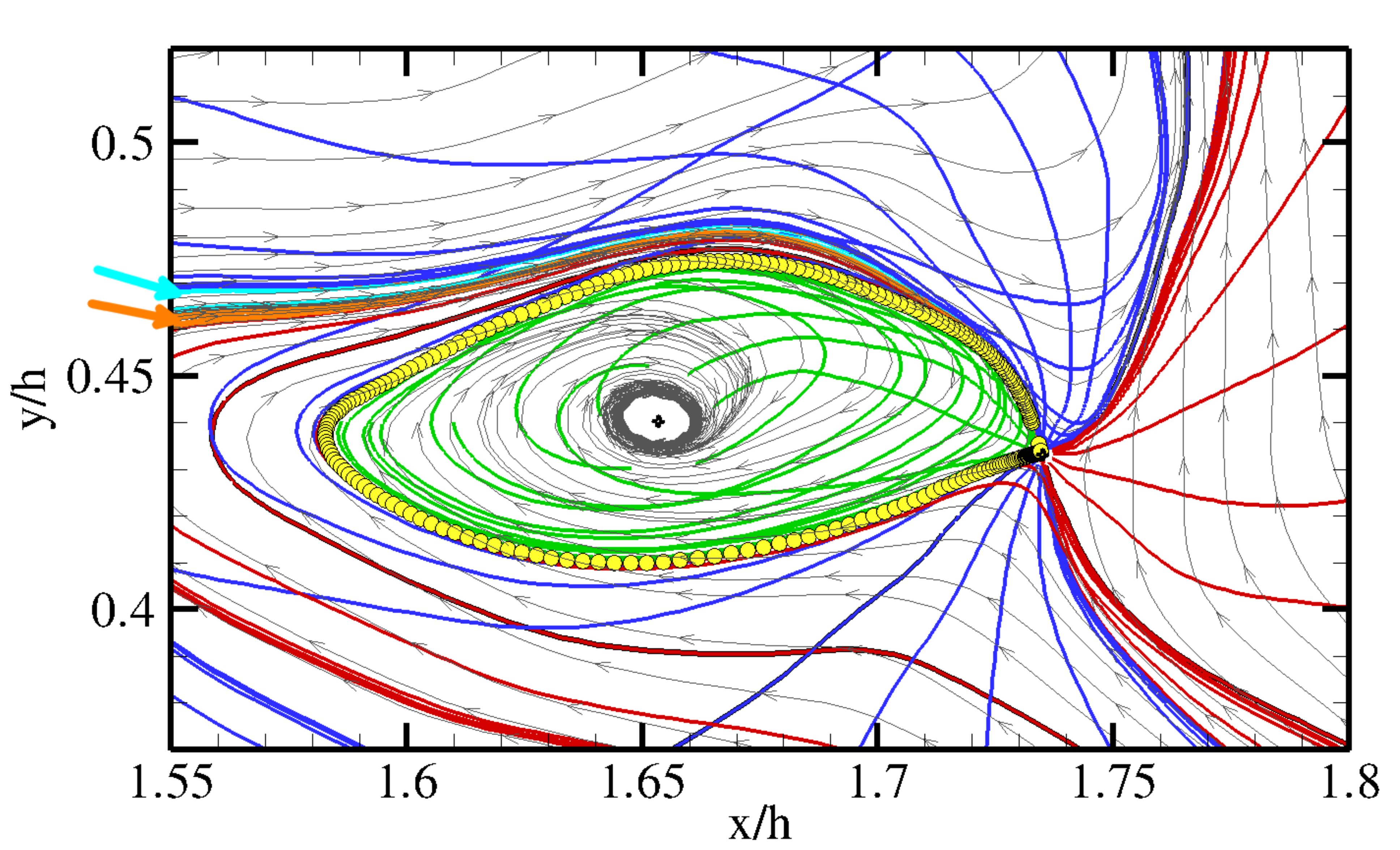}
}    
\caption{\label{fig:VtxInitEnd}
Concept of closed lines $\gamma_\phi$ starting from
a saddle point. (a) illustrates the situation at the beginning of the algorithm, whereas (b) corresponds to the solution at convergence. Streamlines of the vector field are depicted in
light gray. Black  solid lines correspond to the lines
starting in the principal directions of the saddle point.
Colour lines refer to the boundary condition of each line
(blue: exit of velocity field ; green, red, cyan and orange: attraction by a
 critical point). The cyan and orange arrows in (b) highlight two vortex region candidates created by the extraction algorithm. The area materialized by yellow circles corresponds to the vortex region $\Omega$ associated to the closed lines drawn in green.
}
\end{figure*}

\subsection{Particle image velocimetry and vortex detection}
Finally, for evaluating and analyzing the baseline, the best open-loop actuation, 
and the best closed-loop actuation laws (see section \ref{sec:results}), the vector field statistics are obtained with a Stereoscopic PIV acquisition system (LaVision) and its DaVis 7.3 software.
Three-component PIV measurements are taken in the vertical symmetry plane of the downward flow induced  at the mid-plane of one AVG (see Fig.~\ref{fig:RampSketch}).
An Nd:YAG laser (Quantel, EverGreen) generating two pulses of $200\,\mathrm{mJ}$ each at a wavelength of $532\,\mathrm{nm}$ is located above the test section.
A streamwise slit in the test section roof enables the vertical laser light sheet to reach the model.
%The optical setup is chosen to generate a sheet as thin as possible (about $1\,\mathrm{mm}$) in the proximity of the model.
Images are captured with two CCD cameras (Imager LX 11M) mounted on opposite sides of the light sheet to obtain 
forward-scattering. Finally, Scheimpflug adapters are used to obtain a focused area 
despite the viewing angle of about $45$\textdegree. 
The vector fields are computed with a final interrogation window of
$32\times 32$ pixels ($50$\% overlap), giving a grid of
$256$ points in the streamwise direction and $148$ points in the transverse direction.
The space resolution is $2.4\,\mathrm{mm}$ corresponding to $0.024\,x/h$.
The vector field statistics are achieved with the capture of 
$2\,500$ independent vector 
fields acquired at a sampling frequency of $1.6\,\mathrm{Hz}$.
This leads to statistical errors of the mean and second-order 
moments equal to $1$\% and $3$\%, respectively, for a $97$\% 
confidence interval \citep{Benedict1996EiF}.

Instantaneous PIV vector fields are also used to characterize two-dimensional vortex regions.
Many classical algorithms of vortex extraction ($\lambda_2$ and Okubo-Weiss Q criteria for instance)
are based on a scalar indicator function, whose magnitude relates to the strength of vortex activity.
In this framework, the extracted regions depend on a threshold value that is often chosen arbitrarily.
Furthermore, methods based on this approach often do not identify correctly individual vortices and are usually not able to separate adjacent vortices.
Another approach for determining the shape of a vortex is to purely rely on geometrical methods. 
Recently, \citet{Petz2009} proposed an algorithm for vortex region extraction that makes use of vector field topology.
Their definition of a vortex region is based on the generalization of the concept of closed streamlines loops.
For a divergence-free vector field, the union of all closed streamlines defines intuitively a vortex  region.
By extension, the authors generalize the streamline criterion to vector fields with divergence, by introducing closed lines whose tangent present a constant incident angle to the vector field. The union of these lines defines a vortex region candidate that is discriminated by imposing that vortex regions are bounded by closed loops that start and end at saddle points. It can be proved by continuity that at least one saddle point is included in the closure of a vortex region \citep{Petz2009}.

Figure~\ref{fig:VtxInitEnd} illustrates the concept of closed lines, hereafter denoted $\gamma_\phi$, that intersect the flow field at a constant angle $\phi$ and enclose at least one saddle point. The tracing of the closed lines starts from a saddle 
point. The principal directions of the fixed point lead to the discrimination of four distinct regions for the closed lines, two oriented towards the upper and lower boundaries of the flow field, and two oriented toward upstream and downstream vortices. As an illustration, let $\Omega$ be the region oriented toward the core vortex in Fig.~\ref{fig:VtxInitEnd}. Mathematically, this region is defined as $\Omega = \left\lbrace\gamma_\phi\,\mid\,\phi\in\left[\phi_{\mathrm{Min}};\phi_{\mathrm{Max}}\right]\right\rbrace$ where $\phi_{\mathrm{Min}}$ and $\phi_{\mathrm{Max}}$ correspond to the angles specified by the two principal components
designing the region. The lines $\gamma_\phi$ whose angle is close to the mean value of $\left[\phi_{\mathrm{Min}};\phi_{\mathrm{Max}}\right]$ 
describe a straight path from the saddle point to the vortex core, whereas lines associated to angles close to $\phi_{\mathrm{Min}}$ and $\phi_{\mathrm{Max}}$ describe curves looping around the vortex core 
before reaching it (see green lines in Fig.~\ref{fig:VtxInit}).

The algorithm of \citet{Petz2009} presents two steps. First, the critical points 
from a vector field are extracted and sorted by class (saddle 
point, attracting/repelling node, attracting/repelling focus, 
center). The second step consists to trace the lines $\gamma_\phi$ for different values of $\phi$ and then to identify vortex region candidates by clustering neighbouring lines with similar behaviour. For each saddle point, the 
principal directions are determined and a first set of lines $\gamma_\phi$ is generated with a regular distribution of angles $\phi$ between two directions (see Fig.~\ref{fig:VtxInit}). The type of boundary condition of each line (attraction to a critical point or exit of the velocity field) is then determined, 
allowing the creation of vortex region candidates by clustering the lines with the same boundary condition (see 
Fig.~\ref{fig:VtxEnd}). During the algorithm, new regions can be 
discovered and their bounds have also to be determined. This  
iterative process takes end when the boundary lines of two 
neighbouring regions present a difference of angle $\phi$ equal to $\pi / 2^{15}$.
The closed lines $\gamma_\phi$ and the critical points are kept during the 
algorithm. This allows to sort the different vortex regions (keeping one particular vortex region among the set of regions enclosing the same critical points) and to produce a hierarchy of vortex regions (several regions being enclosed in a larger one).

\begin{figure}
  \includegraphics[width=0.45\textwidth]{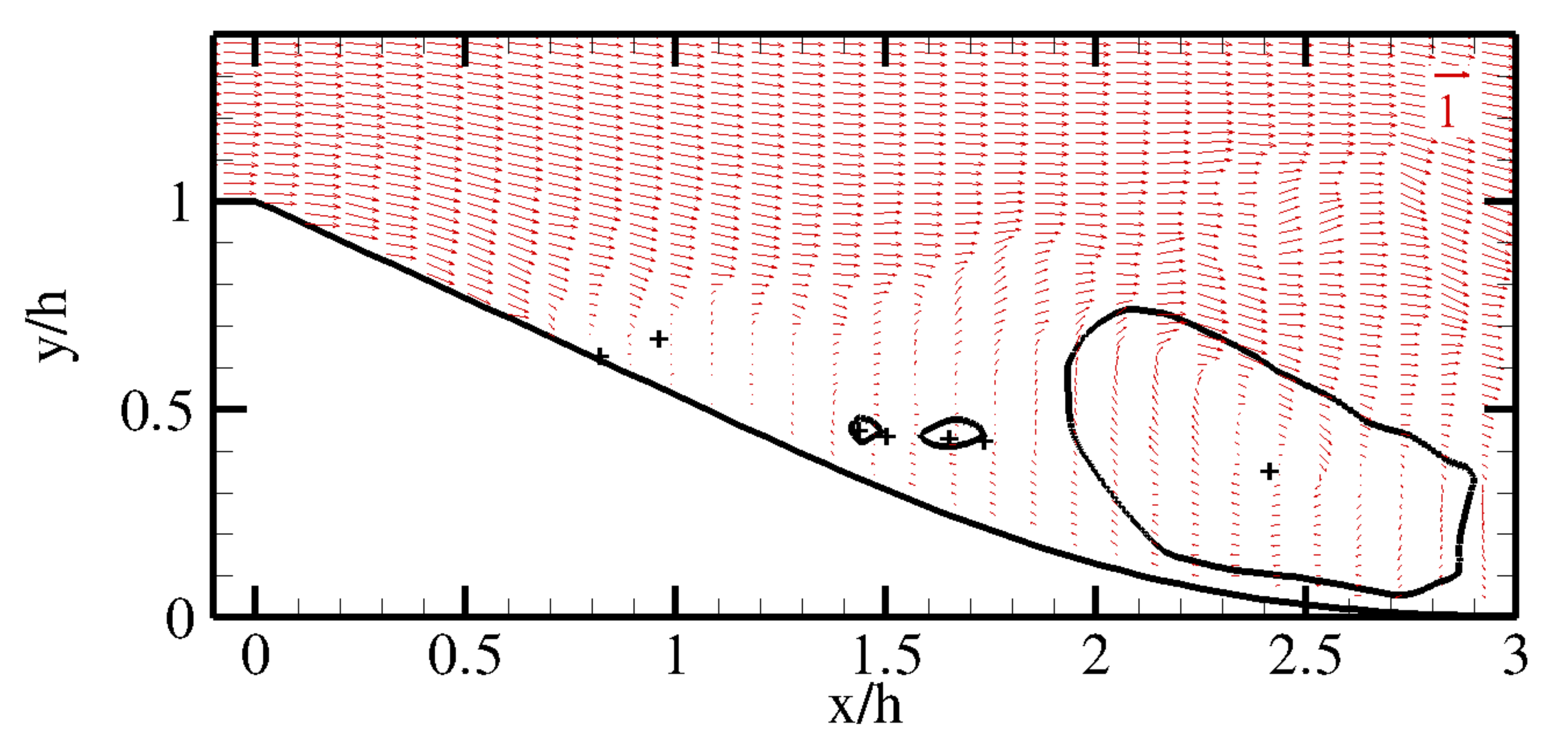}
\caption{
Instantaneous vector field velocity above the ramp and extracted vortex regions (black contours). The black crosses correspond to the detected critical points.
}
\label{fig:VtxVF}
\end{figure}

A typical example of the vortex regions extracted by this algorithm is presented in Fig.~\ref{fig:VtxVF}. This algorithm is used in Sec.~\ref{sec:controlled_flow_analysis} to determine the statistical distribution of the area of vortex regions for the 
different controlled flows.

\section{Genetic Programming \\Control}
\label{sec:MLC}

\subsection{Control Design}
Genetic Programming Control \citep{Parezanovic2015ftac,Gautier2015jfm} is 
a model-free control method designed to determine non-linear control laws 
for a non-linear complex dynamical system in an unsupervised, data driven 
manner. GPC is an evolutionary algorithm largely based on classical  
genetic programming \citep{Koza1992book,Koza1999book} and adapted to 
determine experimental control laws.

\begin{figure}
	\centering
	\begin{overpic}[width=0.9\linewidth]{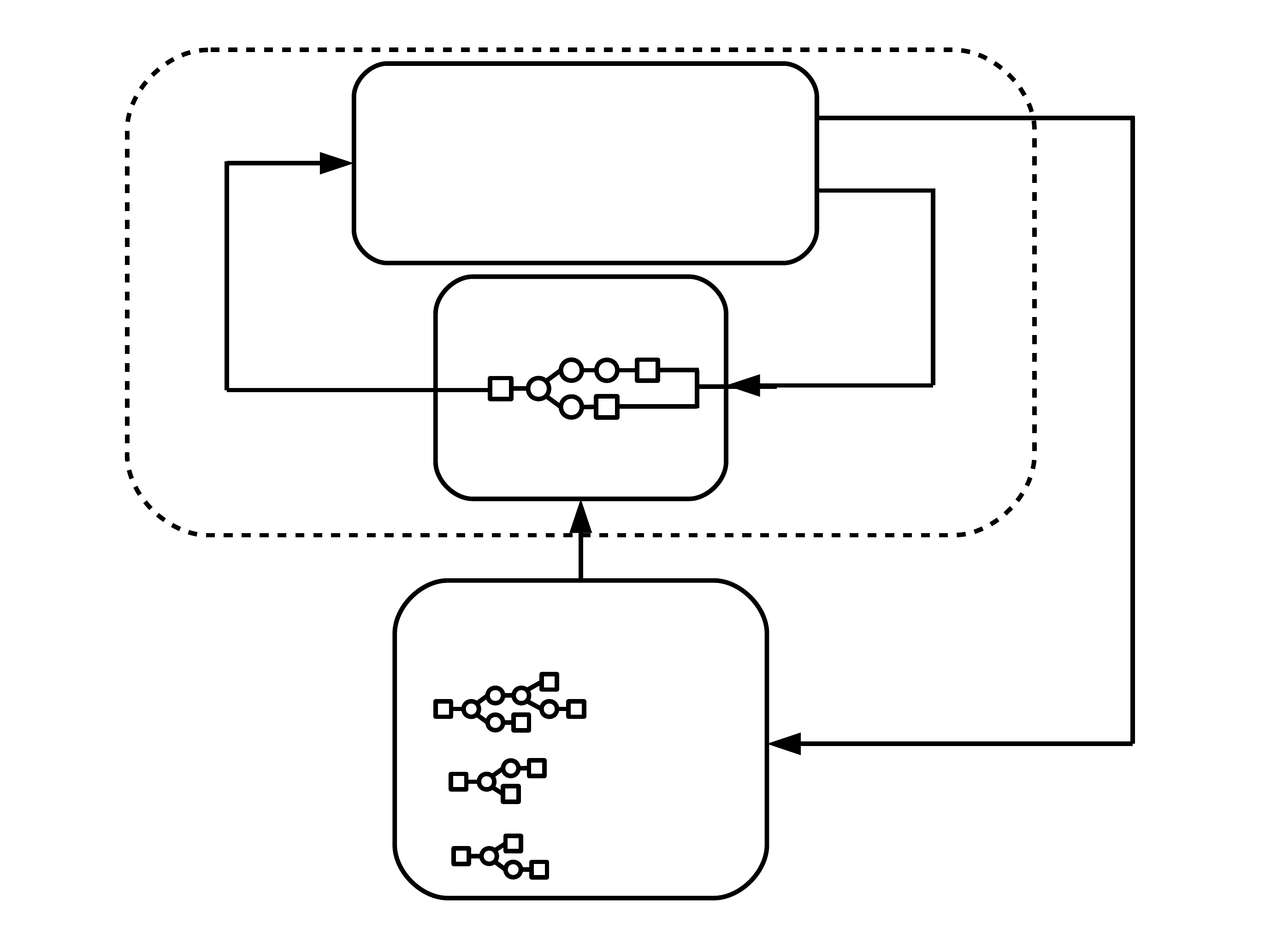}
	\put(25.5,61){\begin{tabular}{rcl}
			$\frac{d\mathbf{a}}{dt}$ &$=$ &$\mathbf{F} \left( \mathbf{a}, \mathbf{b}\right),\,$\\
			$\mathbf{s}       $&$=$&$ \mathbf{H} \left( \mathbf{a}\right)$
		\end{tabular}
	}
		\put(37,48.9){$\mathbf{b}=\mathbf{K} \left(\mathbf{s}\right)$}
		\put(42,25){GPC}	
		\put(12,37){\begin{tabular}{c}
				Real-time\\
				loop (fast)
			\end{tabular}
		}
		\put(61,22){\begin{tabular}{c}
				Learning\\
				loop (slow)
			\end{tabular}
		}
		\put(13,52){$\mathbf{b}$}
		\put(75,52){$\mathbf{s}$}
		\put(91,40){$J_i$}
		\put(45,37){$\mathbf{K}_i$}
		\put(51,18){$J_1$}
		\put(51,6.5){$J_{N}$}
		\put(51.5,12){$\vdots$}
	\end{overpic}
	\caption{General principle of GPC. Two loops are present. The inner loop is the actual control loop which is on a real-time basis. The dynamical system (upper box) feeds the sensors $\mathbf{s}$ to the controller (middle box). The controller uses the control law corresponding to the $i$-th individual ($\mathbf{K}_i$) in order to compute the actuation command $\mathbf{b}$ which is fed back to the dynamical system. The outer loop is the evolutionary learning loop. The GPC algorithm (lower box) is providing the controller with control laws.  After a statistically significant evaluation time, the cost function value $J_i$ is computed for the control law $\mathbf{K}_i$. The GPC algorithm is using the individual values $J_1,\dots,J_N$ to evolve the population of control laws until the control problem is solved.}\label{fig:GPC_scheme}
\end{figure}
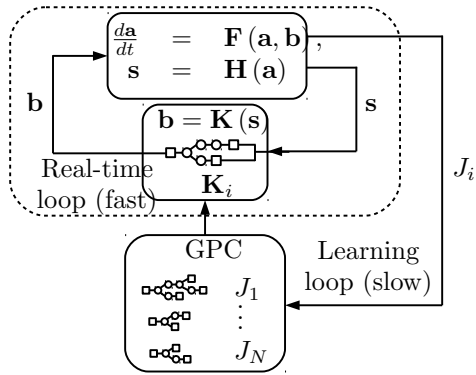

A general view of the algorithm implementation can be seen in 
Fig.~\ref{fig:GPC_scheme}. The experiment is represented by the 
dynamical system:
\begin{equation}
\begin{aligned}
\frac{{\mathrm d}\mathbf a}{\mathrm dt}&=\mathbf F(\mathbf a,\mathbf b),\\
\mathbf{s}       &= \mathbf{H} \left( \mathbf{a}\right),\\
\mathbf{b} &=\mathbf{K(s)}
\end{aligned}
\end{equation}
 with $\mathbf a \in \mathbb{R}^{N_a}$ representing the states, $\mathbf s \in \mathbb{R}^{N_s}$ representing the measurements on the system and $\mathbf b \in \mathbb{R}^{N_b}$ representing the control laws. $\mathbf F$, $\mathbf H$, $\mathbf K$ are respectively the evolution operator, the measurement function and the control laws. The control problem is defined by a cost function $J(\mathbf{s},\mathbf{b})$ to be minimized. In the GPC framework, the control laws are represented by expression trees, containing arbitrarily complex combinations of user defined functions, operations, sensors and random constants. The learning GPC process is used to determine the control law best fitted to be used in the inner real-time feedback loop. Once the best control law is determined, the learning loop can be disconnected.
 
\begin{figure}
 	\begin{center}
 		\includegraphics[width=.35\textwidth]{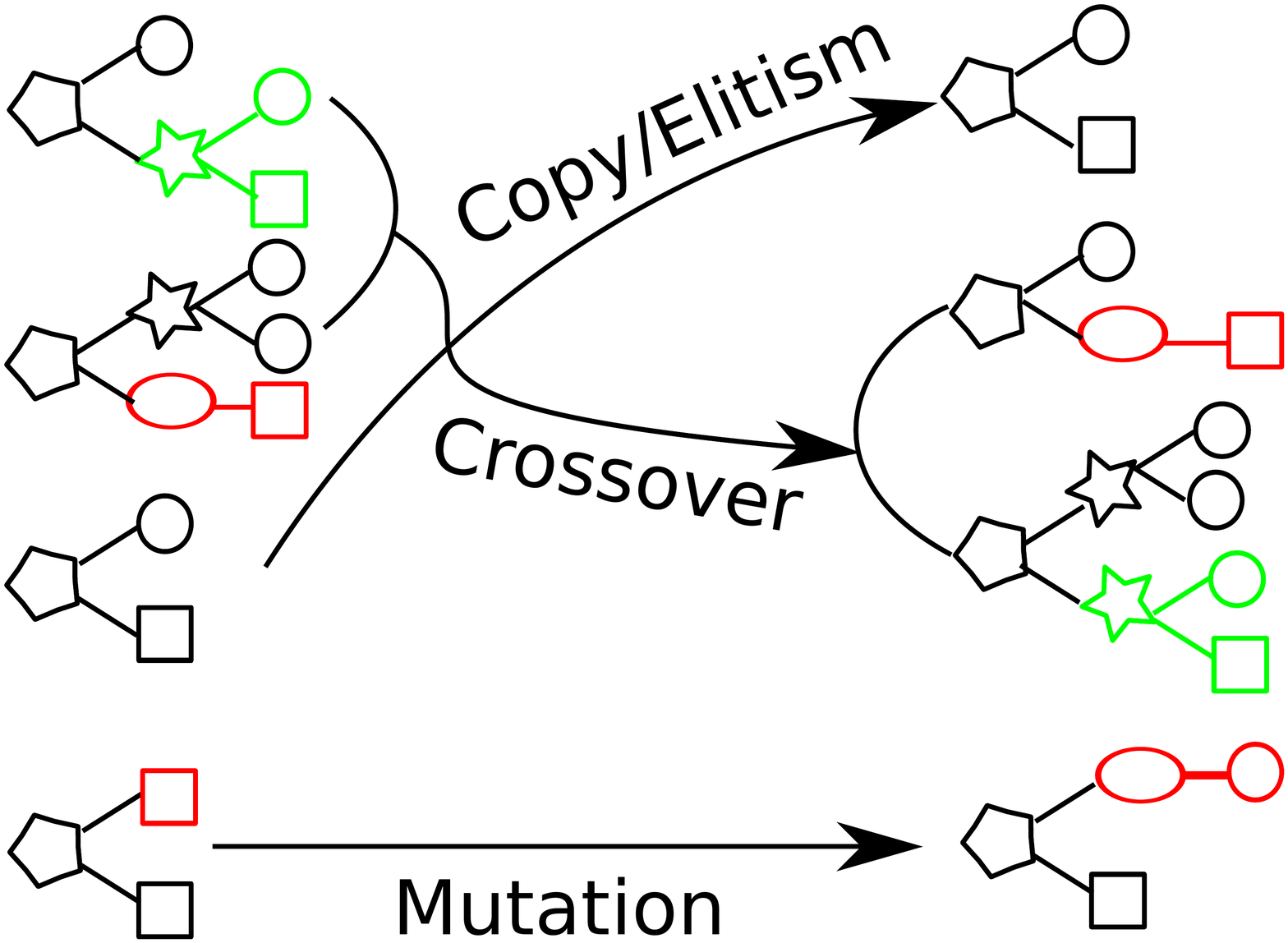}
 	\end{center}
 	\caption{Graphical illustration of the different genetic operations: copy,
 		crossover, mutation.}
 	\label{fig:gen_operations}       % Give a unique label
 \end{figure}

The learning process is achieved as follows. A first population of $N$ 
individuals representing the control laws $\mathbf{K}_i$ with 
$i\in[1,\dots,N]$ is randomly created. Each of these individuals is 
tested in the real-time dynamical system loop during the evaluation time 
$T$. At the end of the evaluation time, a cost function value $J_i$ is 
given to each individual $\mathbf{K}_i$. A selection process determines 
which individuals are chosen for the population evolution. For each of the 
$N$ new individuals to produce, a tournament is achieved between randomly 
chosen evaluated individuals, the individual with the lowest cost function 
value being selected for evolution. The selected individuals then go 
through genetic operations: replication (copy of the selected individual 
to the next generation), mutation (partial random alteration of the 
content of the individual) and crossover (partial exchange of the content 
of two selected individuals). Also, the five best individuals of the 
evaluated population are directly copied into the next generation in order 
to ensure that the following generation is at least as good as the 
preceding one. These operations are illustrated in 
Fig.~\ref{fig:gen_operations}.
 
Additionally to this classical genetic programming implementation, GPC 
encompasses modifications to account for the experimental evaluation of 
the individuals: when an individual appears again in a new generation, it 
is further evaluated and its cost function value is averaged. Also, the 
three best individuals of each generation is reevaluated three times. 
This ensures that the convergence process is not dramatically affected by 
measurement noise and errors by eliminating intrinsically bad individuals 
that were accidentally assigned a low cost function value. Also it 
increases the statistical significance of the best individual's cost 
function value which allows to reduce the impact of having a suboptimal 
evaluation time $T$ and then reduce the overall process time cost. 
 
Once a new generation is created, a new evaluation of the whole population 
can be achieved. The learning process is stopped either when the cost 
function value of the best individual reaches a known global minimum 
(theoretical optimality stop criteria, though this is unlikely in an experiment), when no improvement has been achieved through 
several generations (empirical optimality stop criteria), when the minimal cost function value reaches a 
user predetermined value that reflects an acceptable performance (performance stop criteria) or when a prescribed number of generations has been reached (operational cost stop criteria). In those 
cases the best individual is used in the real-time controller and the 
learning loop can be disconnected.
 
\subsection{Experimental Implementation}

In the experiments, the GPC learning loop is implemented on a standard 
computer in Matlab, whereas the inner loop is implemented on an Arduino 
microcontroller. This Arduino acquires the experimental sensors signal and 
computes the control command sent to the actuators according to the 
current individual in the real-time closed-loop control. For each 
generation, the outer loop generates expressions for the individuals, 
compiles and uploads them on the inner-loop Arduino. The Arduino then 
loops between the evaluation of each control law $\mathbf{K}_i$ during a 
time $T=5\,\text{s}$, with a $3\,\text{s}$ resting time in between, and 
returns the cost function values $J_i$ to the learning loop. 

The mean values of the hot-film sensors for the baseline flow ($\mathrm{HF}_{i,0}$) and 
for continuous blowing of the jets ($\mathrm{HF}_{i,\mathrm{Blow}}$)  
are used at the beginning of each 
generation and every 100 evaluations of the cost functions to normalize the hot-film sensors.

For each of the two hot-film sensors, three virtual sensors are created 
which leads in total to $N_s=6$. 

The first virtual sensor is calibrated to the average blowing and not blowing
values, \textit{i.e.}
\begin{equation}
s_i = \frac{\langle \mathrm{HF}_i \rangle - \mathrm{HF}_{i,0}}{\mathrm{HF}_{i,\mathrm{Blow}}-\mathrm{HF}_{i,0}}, \qquad i\in \{0,1\},
\label{eq:sensors1}
\end{equation}
where $\langle\cdot\rangle$ denotes a moving average over ten measurements.

The second one is the instantaneous calibrated value of the sensor:
\begin{equation}
s_{i+2} = \frac{\mathrm{HF}_i - \mathrm{HF}_{i,0}}{\mathrm{HF}_{i,\mathrm{Blow}}-\mathrm{HF}_{i,0}}, \qquad i\in \{0,1\}.
\label{eq:sensors2}
\end{equation}
The calibration ensures that the virtual sensor values are close to the $[0,1]$ interval.
The third sensor is an instantaneous fluctuation calculated as:

\begin{align}
s_{i+4} & = s_{i+2} -s_{i} \nonumber\\
        & = \frac{\mathrm{HF}_i - \langle \mathrm{HF}_i \rangle}{\mathrm{HF}_{i,\mathrm{Blow}}-\mathrm{HF}_{i,0}} \qquad i\in \{0,1\}.
\label{eq:sensors3}
\end{align}

The operators used to create the individuals are $+$, $-$, $\times$, $/$, 
$\sin$, $\cos$, $\ln$, $\exp$ and $\tanh$. Sensitive operations such 
as $/$ and $\ln$ are protected so that any value in $\mathbb{R}$ can be 
used as arguments. The absolute value of the denominator of $/$ is 
saturated to $10^{-2}$. Also, $x\mapsto\ln(x)$ is modified to 
$x\mapsto\ln(|x|)$ for $|x|>10^{-2}$ and $x\mapsto \ln(10^{-2})$ otherwise. 
Finally, the output of the constructed control laws is passed through the 
Heavyside function to transform the continuous output from $\mathbf{K}_i$ 
to an on/off signal. If $b>0$ (respectively $b<0$), the control is on 
(respectively off).

The physical objective of the control problem is to re-attach the boundary layer in an energy efficient manner. The state of the boundary layer is assessed by the mean wall-friction measured by the hot-films and the average pressure distribution measured by the pressure taps. The actuation cost is assessed by the calculation of the average $c_\mu$ obtained from the flow meter on the pressure tank. This problem corresponding to a multi-objective optimization, the cost function used for GPC is defined as:
 \begin{equation}
 J = J_{\mathrm{HF}} +  \ell_{\mathrm P_{\mathrm{Stat}}} J_{\mathrm P_{\mathrm{Stat}}} + \ell_{\mathrm{Act}} J_{\mathrm{Act}}
 \label{eq:J_final}
 \end{equation}
where $J_i$, $i\in\{\mathrm{HF},\mathrm P_{\mathrm{Stat}},\mathrm{Act}\}$, corresponds to the cost functions for the different criteria to optimize and where $\ell_{\mathrm P_{\mathrm{Stat}}}$ and $\ell_{\mathrm{Act}}$ are penalization coefficients.\\
 
The term $J_{\mathrm{HF}}$ related to the wall friction is calculated using:
 \begin{equation}
 J_{\mathrm{HF}} = 
 \frac{1}{N_{\mathrm{HF}}} 
 \sum_{i=1}^{N_{\mathrm{HF}}} 
 \left[1-
 \tanh\left(\frac{\langle {\mathrm{HF}}_i\rangle}{{\mathrm{HF}}_{i,0}}-1\right)\right]
 \label{eq:J_HF}
 \end{equation}
 where $N_{\mathrm{HF}}$ is the number of hot-film sensors.
 
 The term $J_{\mathrm P_{\mathrm{Stat}}}$ related to the static pressure sensors is
 obtained using:
 \begin{equation}
 J_{\mathrm P_{\mathrm{Stat}}} = 
 \int_T\,\frac{1}{
 \displaystyle 0.1 + \sum_{i=1}^{N_{\mathrm{P}}} (P(x_i)-P_0(x_i))^2 \frac{x_{\mathrm{max}}-x_i}{x_{\mathrm{max}}-x|_{x=0}}}\,{\mathrm d}t, 
 \label{eq:J_p}
 \end{equation}
where $N_{\mathrm{P}}$ is the number of static pressure sensors used for the estimation and $x_i$ their streamwise coordinates.
In the experiments, the estimation of $J_{\mathrm P_{\mathrm{Stat}}}$ is based only on the pressure sensors located from the sharp edge ($x=0$) to $x_{\mathrm{max}}=4.7\,h$. In addition, $P_0$ is the baseline pressure, and $0.1$ in the denominator is added to
 prevent the division by zero. 
 The $J_{\mathrm P_{\mathrm{Stat}}}$ cost function penalizes an attachment point located far 
 from the sharp edge and therefore large separation bubble as the separation
 point is fixed by the geometric discontinuity. 
 
The actuation penalization term $J_{\mathrm{Act}}$ is obtained using:
 \begin{equation}
 J_{\mathrm{Act}}=\int_T\,\frac{c_{\mu}}{V_{\mathrm{Jet}}}\,{\mathrm d}t.
 \label{eq:J_act}
 \end{equation}

In Sec.~\ref{sec:results}, we will keep constant the weight $\ell_{\mathrm P_{\mathrm{Stat}}}$ and analyse the influence of $\ell_{\mathrm{Act}}$ on the best GPC actuation laws.
\section{Results}
\label{sec:results}
The presentation of the results will be done in three steps. First, in Sec.~\ref{sec:online_results}, we present the GPC results through the data that are used during the Genetic Programming runs, \textit{i.e.} the signal of the HF sensors, the pressure sensors, and the estimation of the momentum coefficient $c_\mu$. Then, in Sec.~\ref{sec:offline_eval}, we perform an a posteriori analysis of the previous GPC results with the PIV measurements obtained offline. Finally, an in-depth analysis of a particular actuation law (individual 7), ranked among the most efficient by GPC, is done in Sec.~\ref{sec:controlled_flow_analysis}. In the following, the use of 
individual and control or actuation law is strictly equivalent.

\subsection{GPC Results}
%\subsection{Online Results}
\label{sec:online_results}
During the experimental determination of the closed-loop control laws, 
the use of the HF sensors is twofold: as an input signal for the computation of the control law, and to estimate $J_{\mathrm{HF}}$ according to \eqref{eq:J_HF}.
Both calculations are implemented in real time on the Arduino. The
pressure data on the ramp is collected on a different computer and used to compute $J_{\mathrm P_{\mathrm{Stat}}}$ according to \eqref{eq:J_p}. These results are weighted with the estimation of $J_{\mathrm{Act}}$ to determine $J$ needed to rank the individuals in the GPC algorithm.

Three different runs of genetic programming control are performed. 
For all the runs, the weight $\ell_{\mathrm P_{\mathrm{Stat}}}=\tfrac{1}{30}$ is 
kept constant. In addition, the probability of crossover, copy and 
mutation are fixed to $p_{\mathrm{Cross}}=0.7$, $p_{\mathrm{Copy}}=0.1$ and  
$p_{\mathrm{Mut}}=0.2$, respectively. In order to test the influence of the amount of actuation on the results, 
three different values of penalization weights $\ell_{\mathrm{Act}}$ are considered, namely $2.5, 0.8$ and $0.6$. $\ell_{\mathrm{Act}}=2.5$ corresponds to a strong penalization whereas setting  $\ell_{\mathrm{Act}}=0.8$ and  $\ell_{\mathrm{Act}}=0.6$ lower the level of penalization but still prevent constant blowing as a solution.  
\begin{figure}[h]
\includegraphics[width=0.45\textwidth]{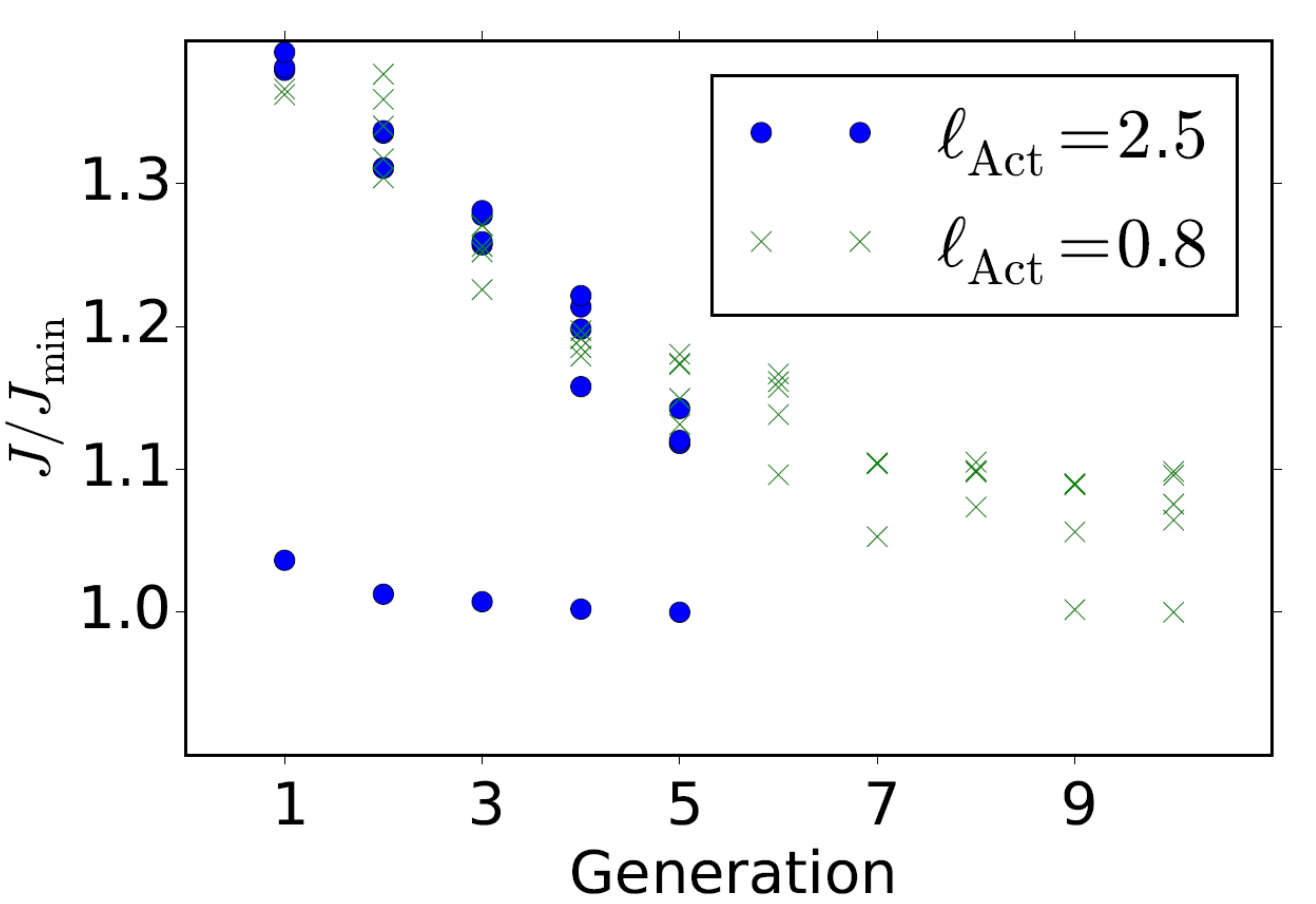}
\caption{Variation of the normalized cost function value $J$ over the generations for two GPC runs, one of 5 generations with $\ell_{\mathrm{Act}}=2.5$ and one of 10 generations with $\ell_{\mathrm{Act}}=0.8$.The five best individuals of each run are plotted. Even when the first generation contains already effective individuals (case  $\ell_{\mathrm{Act}}=2.5$), we can note an overall improvement of the high ranking individuals of the population with $20$ to $30\%$ reduction of $J$ in few generations. This demonstrates how the population is stirred towards the best individual to explore local minima and further improve the best result.
}
\label{fig:evol_of_J}
\end{figure}
Recently, \citet{Debien2015TSFP9} showed that for the same configuration of
sharp edge ramp the best open-loop actuation has a frequency
$f_{\mathrm{Pulse}}=30\,\text{Hz}$ and a duty cycle of 50\%. 
In the following,
we benchmark against this best periodic forcing eight particular actuation laws, namely:
i) individuals 1 and 2 are taken out the first GPC run ($\ell_{\mathrm{Act}}=2.5$) after 5 generations, individual 1 being the best individual of all generations ; 
ii) individuals 3, 4, and 5 are the three best actuation laws obtained from the second GPC run ($\ell_{\mathrm{Act}}=0.8$) after 10 generations ; 
iii) individuals 6, 7, and 8 are obtained from the third GPC run ($\ell_{\mathrm{Act}}=0.6$) after 5 generations, individual 6 being the best individual of all generations.

The evolution of the normalized cost function $J$ over the generations is shown in Fig.~\ref{fig:evol_of_J} for two different values of $\ell_{\mathrm{Act}}$. For 
each generation, the $J$ value of the five best performing 
actuation laws of each GPC run
%over the eight discussed in this paper 
is plotted to show not only the improvement of 
the most effective control law but also to illustrate the global convergence of the population toward effective control laws for an increasing number of generations.
One can see that with an increase of the number of generations, the 
performance of the best actuation would likely have increased. Due to 
experimental time constraints, the number of generations is kept low. 

The $J$ values for the different individuals are recorded in Tab.~\ref{tab:Js}. 
These values are given, once with the specific value of $\ell_{\mathrm{Act}}$ used during
the GPC run, and once with an arbitrarily chosen value ($\ell_{\mathrm{Act}}=0.5$) to have 
a fair comparison with the best open-loop solution which has been chosen to achieve maximum 
performance regardless of actuation cost. The cost corresponding to each control problem 
defined by a different $\ell_{\mathrm{Act}}$ value has also been computed for the best 
open-loop.
\begin{table}
\caption{Performance comparison of the different actuation laws based on
  their $J$ values. 
  The penalization coefficients
  between the different terms of the $J$ function
  were applied once with $\ell_{\mathrm{Act}}=\ell_{\mathrm{Act}}^{\text{Run}}$ kept at the 
  value employed during the GP run and once with $\ell_{\mathrm{Act}}=0.5$ for
  comparison with the best open-loop control. The right column introduces the reference name of the
GPC law in the manuscript. For details, see text.
}
\label{tab:Js}       % Give a unique 
\centering
\begin{tabular}{llll}
	\hline\noalign{\smallskip}
	& $J|_{\ell_{\mathrm{Act}}=\ell_{\mathrm{Act}}^{\text{Run}}}$  
        & $J|_{\ell_{\mathrm{Act}}=0.5}$ 
        & Reference\\
	\noalign{\smallskip}\hline\noalign{\smallskip}
	\multicolumn{3}{c}{$\ell_{\mathrm{Act}}^{\text{Run}}=2.5$}\\\\
%	\noalign{\smallskip}\hline\noalign{\smallskip}
	Best open-loop  & 1.14 &  0.38 & \\
	Best GPC law     & 0.88 &  0.38 & Indiv. 1\\
	Subopt. GPC law  & 1.12 &  1.00 & Indiv. 2\\
	\noalign{\smallskip}\hline\noalign{\smallskip}
	\multicolumn{3}{c}{$\ell_{\mathrm{Act}}^{\text{Run}}=0.8$}\\\\
%	\noalign{\smallskip}\hline\noalign{\smallskip}
    Best open-loop  & 0.50 &  0.38\\
	 Best GPC law & 0.45 &  0.40 & Indiv. 3\\
	 Subopt. GPC law  & 0.47 &  0.42& Indiv. 4\\
 	 Subopt. GPC law  & 0.52 &  0.41& Indiv. 5\\
	\noalign{\smallskip}\hline\noalign{\smallskip}
	\multicolumn{3}{c}{$\ell_{\mathrm{Act}}^{\text{Run}}=0.6$}\\\\
%	\noalign{\smallskip}\hline\noalign{\smallskip}
    Best open-loop  & 0.42 &  0.38\\
	Best GPC law  & 0.42 &  0.39 &Indiv. 6 \\
	Subopt. GPC law  & 0.45 &  0.41 & Indiv. 7\\
	 Subopt. GPC law  & 0.46 &  0.45 & Indiv. 8\\
	\noalign{\smallskip}\hline
\end{tabular}
\end{table}

This table shows how GPC consistently finds closed-loop control laws performing better or similarly (where pure performance is predominant) to the best open-loop control as computed through the cost function used for their respective determination. Interestingly, at the exception of individual 2, all closed-loop control laws perform reasonably well with a cost function adjusted for low actuation cost ($\ell_{\mathrm{Act}}=0.5$). As a matter of fact, the  $J|_{\ell_{\mathrm{Act}}=0.5}$ values seem to indicate that all these individuals are roughly interchangeable. For instance individual 1 obtained with a high actuation penalization, individual 7 obtained with a low actuation penalization and the best open-loop show similar values. Nonetheless, the decomposition of the cost function in each of its constituent (see Tab.~\ref{tab:J_indivs}) shows that the final value of $J$ is obtained through different mechanisms for each GPC determined individual: some focus on performance, others on economy.

\begin{table}
\caption{Performance comparison of the different actuation laws based on their
  $J_i$ ($i\in\{\mathrm{HF},\mathrm P_{\mathrm{Stat}},\mathrm{Act}\}$) values.
  The weighting coefficients $\ell_i$ are not applied.
}
\label{tab:J_indivs}       % Give a unique label
\centering
\begin{tabular}{cllll}
\hline\noalign{\smallskip}
 & $J_{\mathrm{HF}}$ & $J_{\mathrm P_{\mathrm{Stat}}}$ & $J_{\mathrm{Act}}$ \\
\noalign{\smallskip}\hline\noalign{\smallskip}
Best open-loop & 0.04  & 4.6 & 0.38\\
Indiv. 1 & 0.05  & 6.2 & 0.25\\
Indiv. 2 & 0.11  & 25.8 & 0.06\\
Indiv. 3 & 0.06  & 7.5 & 0.18\\
Indiv. 4 & 0.07  & 7.7 & 0.18\\
Indiv. 5 & 0.06  & 5.1 & 0.36\\
Indiv. 6 & 0.06  & 5.7 & 0.28\\
Indiv. 7 & 0.06  & 4.5 & 0.40\\
Indiv. 8 & 0.08  & 8.8 & 0.15\\
\noalign{\smallskip}\hline
\end{tabular}
\end{table}

\begin{figure}[!t]
  \includegraphics[width=0.45\textwidth]{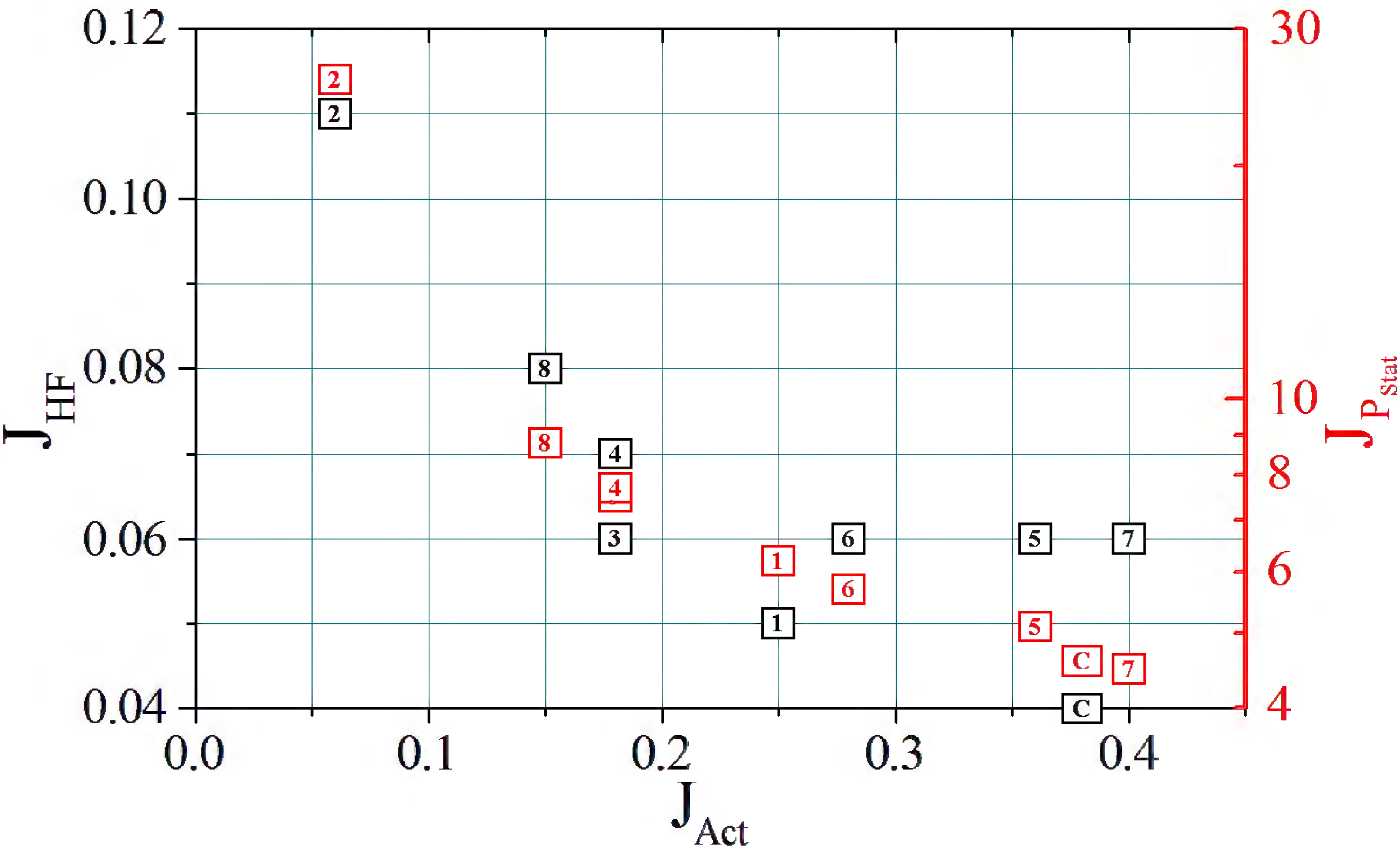}
\caption{Variation of the cost function values $J_{\mathrm HF}$ and $J_{\mathrm 
P_{\mathrm{Stat}}}$ versus $J_{\mathrm{Act}}$ for the GPC individuals and the best open-loop solution (label c).
}
\label{fig:Jfunc}
\end{figure}

This performance-actuation trade-off is further illustrated in 
Fig.~\ref{fig:Jfunc} where  $J_{\mathrm HF}$ and
$J_{\mathrm P_{\mathrm{Stat}}}$ are plotted versus $J_{\mathrm{Act}}$. We observe that
$J_{\mathrm{HF}}$ and $J_{\mathrm P_{\mathrm{Stat}}}$ vary as an hyperbolic
function with respect to $J_{\mathrm{Act}}$, even though a noticeable scatter
can be seen for high values of the actuation penalty. This hyperbolic behaviour
reflects the compromise between the cost of the control and its effectiveness.
The set of solutions obtained by GPC is then equivalent to the Pareto frontier
classically used in multi-disciplinary optimization. 
\begin{figure}[t]
  \includegraphics[width=0.45\textwidth]{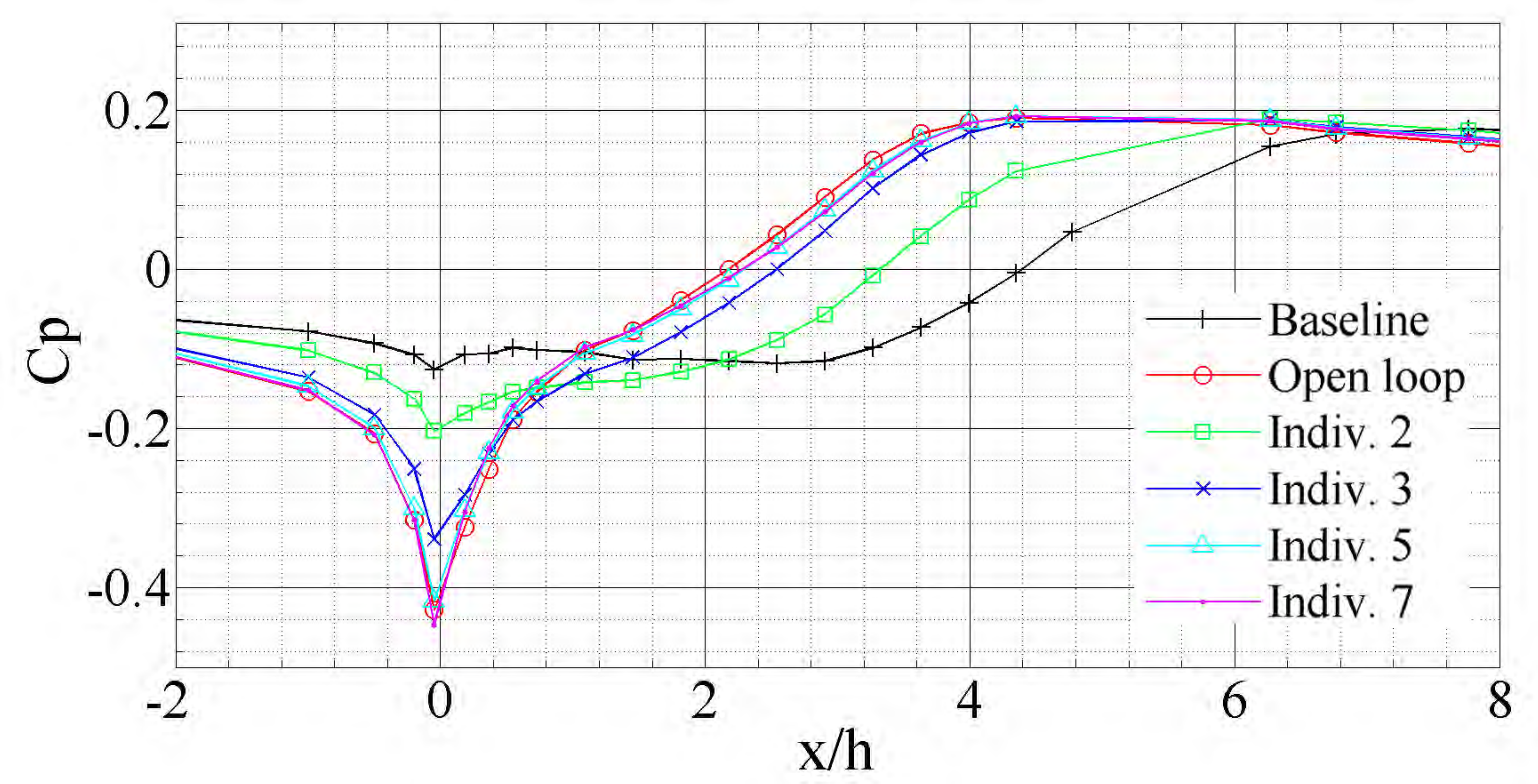}
\caption{Pressure coefficient distribution over the ramp for the baseline, the best open-loop and some GPC individuals.
Only the pressure sensors located in the range $0 \le x/h \le 4.70$ are used for estimating $J_{\mathrm P_{\mathrm{Stat}}}$.
}
\label{fig:Pstat}
\end{figure}
Due to the relatively large spacing of the pressure sensors, the determination of the
position of the reattachment point using pressure sensors can not be
inferred with accuracy and is therefore not clearly visible in 
the $J_{\mathrm P_{\mathrm{Stat}}}$ 
distribution. The evolution of the pressure coefficient ($C_p$) 
over the length of the ramp is presented in
Fig.~\ref{fig:Pstat}. Upstream of the sharp edge ($x/h = 0$),
$C_p$ gradually decreases
towards the ramp due to a favourable pressure gradient induced 
by the proximity of the ramp and by the blowing of the AVGs.
Downstream of the sharp edge of the ramp, the pressure rises up 
to its maximal value ($C_p \approx 0.2$) around
the location of the reattachment point. Further downstream, the pressure
converges towards $C_p \approx 0.19$. 
\begin{figure}[t]
  \includegraphics[width=0.45\textwidth]{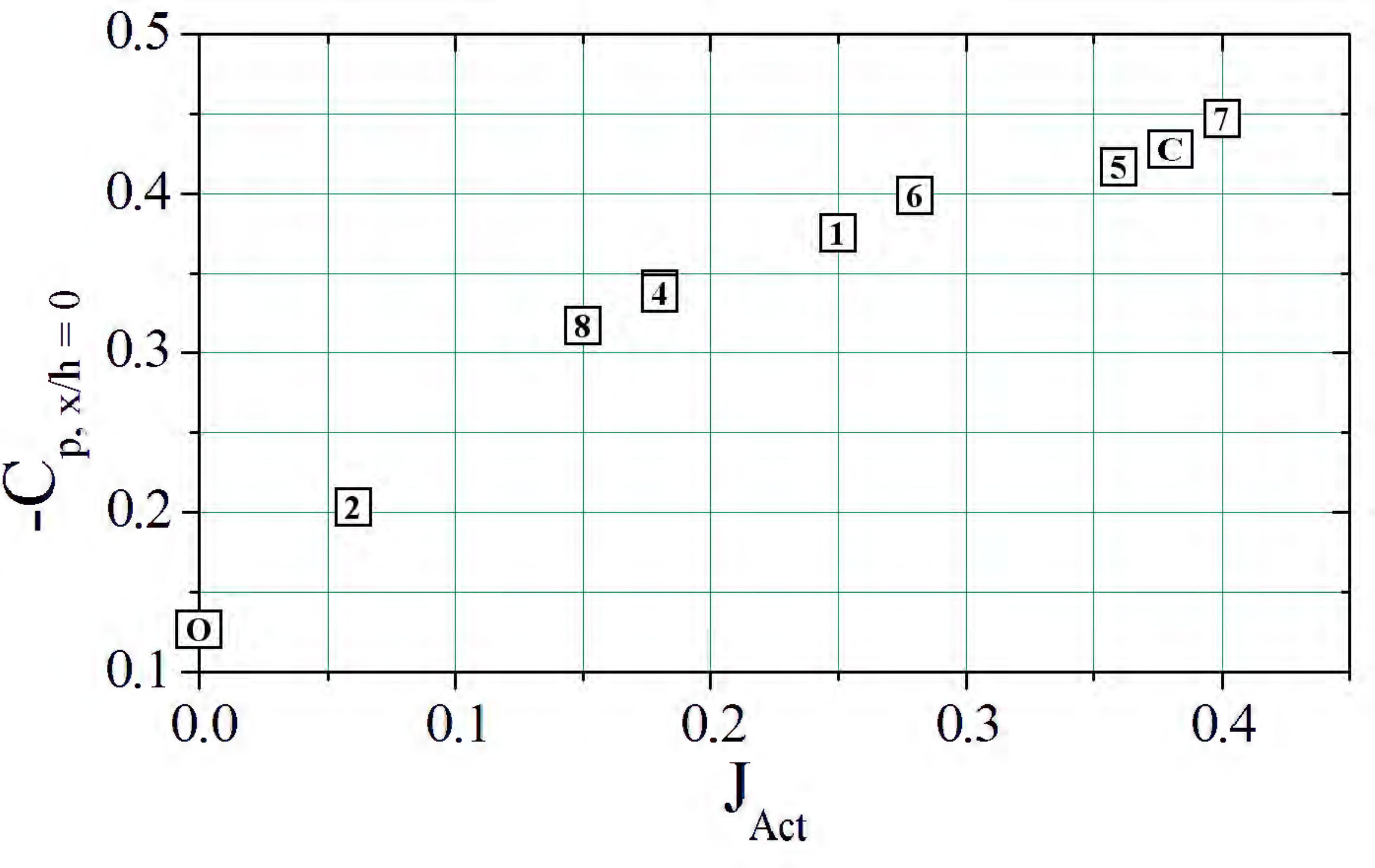}
\caption{Pressure coefficient at the sharp edge ($x/h = 0$) versus $J_{\mathrm{Act}}$ 
for the baseline (label o), the best open-loop (label c) and the GPC individuals.
}
\label{fig:Cpx_Jact}
\end{figure}
Figure~\ref{fig:Pstat} illustrates a direct relationship between actuation cost $J_{\mathrm{Act}}$ (as reported in Tab.~\ref{tab:J_indivs}) and separation control:
the larger $J_{\mathrm{Act}}$ is, the smaller the separation length is.
The $C_p$ distribution measured for the GPC individuals approach to the one obtained for the best open-loop controller as $J_{\mathrm{Act}}$ increases.
Whereas
the pressure distribution around the attachment point provides a direct 
measurement of the effectiveness of the actuation, the pressure at the 
separation point (see Fig.~\ref{fig:Cpx_Jact}) depends mainly on the momentum
coefficient of the AVGs
jets which generates a pressure minimum around the actuators.

Based on these criteria, four individuals seem to yield particular good 
results: individuals 1, 5, 6 and 7. Individual 5 and 7 
present a $J_{\mathrm{Act}}$ value close to the best open-loop case whereas 
individuals 1 and 6 present a 34\% and a 26\% $J_{\mathrm{Act}}$ 
reduction, respectively, but also worse $J_{\mathrm P_{\mathrm{Stat}}}$ 
values (see Fig.~\ref{fig:Jfunc}).

\subsection{Analysis based on PIV measurements} 
\label{sec:offline_eval}
Though the stereo PIV equipment is not able to feed its results in real-time in order to be used in the GPC process, it can be used to assess the controlled flow using the controllers obtained by GPC. In this section we provide an in-depth a posteriori analysis of the controlled flows obtained by GPC, with a detailed study of the actuation mechanism and evaluation from PIV of the flow features. A special focus is done on the individual 7 which appears to be the most interesting.

\subsubsection{AVGs' characteristics}

To evaluate the performance of the different control laws, the AVGs actuation
characteristics are first examined. For the different control laws, the blowing 
velocity $V_{Jet}$ of the AVGs and an estimated ``duty cycle" are reported in 
Tab.~\ref{tab:VGA_charact}. The blowing velocity is computed from a 
calibration law where the input is the pressure of the supplying tank. 
The ``duty cycle" corresponds to the 
ratio between the AVGs' blowing duration and the total acquisition time. Using
these values, the momentum coefficient $c_\mu$ and the energy flow rate
of the blowing jets defined as
\begin{equation}
M_\mathrm{Jet} = \rho_\mathrm{Jet}S_\mathrm{Jet} D_c V_\mathrm{Jet}^3 
  \label{eq:Mj}
\end{equation}
are computed.

Since blowing velocity is kept constant throughout experiments, the cost
function $J_{\mathrm{Act}}$ mostly depends on $D_c$. However, it is worth
noting that $V_{\mathrm{Jet}}$ may vary within a 5--7\% range around its
designed operating value. This occurs in high actuation frequency range
($f_{\mathrm{Pulse}} \geq 250$\,Hz) due to the limited time response of the valves.
For the closed-loop cases, $D_c$ is in the range between 0.07 and 0.49.

\begin{table}[t]
\caption{Characteristics of AVGs during the control phase.}
\label{tab:VGA_charact}       % Give a unique label
\centering
\begin{tabular}{cllll}
\hline\noalign{\smallskip}
 & $V_{\textrm Jet}$ & $D_c$ & $c_\mu \times 10^{-4}$ & $M_\mathrm{Jet}$ \\
\noalign{\smallskip}\hline\noalign{\smallskip}
Best open-loop & 62.3 & 0.50 & 16.5 & 18.0 \\
Indiv. 1 & 62.6 & 0.25 & 8.4 & 9.3\\
Indiv. 2 & 58.1 & 0.07 & 2.0 & 2.05\\
Indiv. 3 & 56.5 & 0.18 & 5.0 & 4.9\\
Indiv. 4 & 56.5 & 0.19 & 5.1 & 5.1\\
Indiv. 5 & 58.3 & 0.46 & 13.3 & 13.6\\
Indiv. 6 & 55.4 & 0.27 & 7.0 & 6.8\\
Indiv. 7 & 57.3 & 0.49 & 13.7 & 13.7\\
Indiv. 8 & 56.7 & 0.18 & 4.8 & 4.8\\
\noalign{\smallskip}\hline
\end{tabular}
\end{table}
\begin{figure}[!h]
  \includegraphics[width=0.45\textwidth]{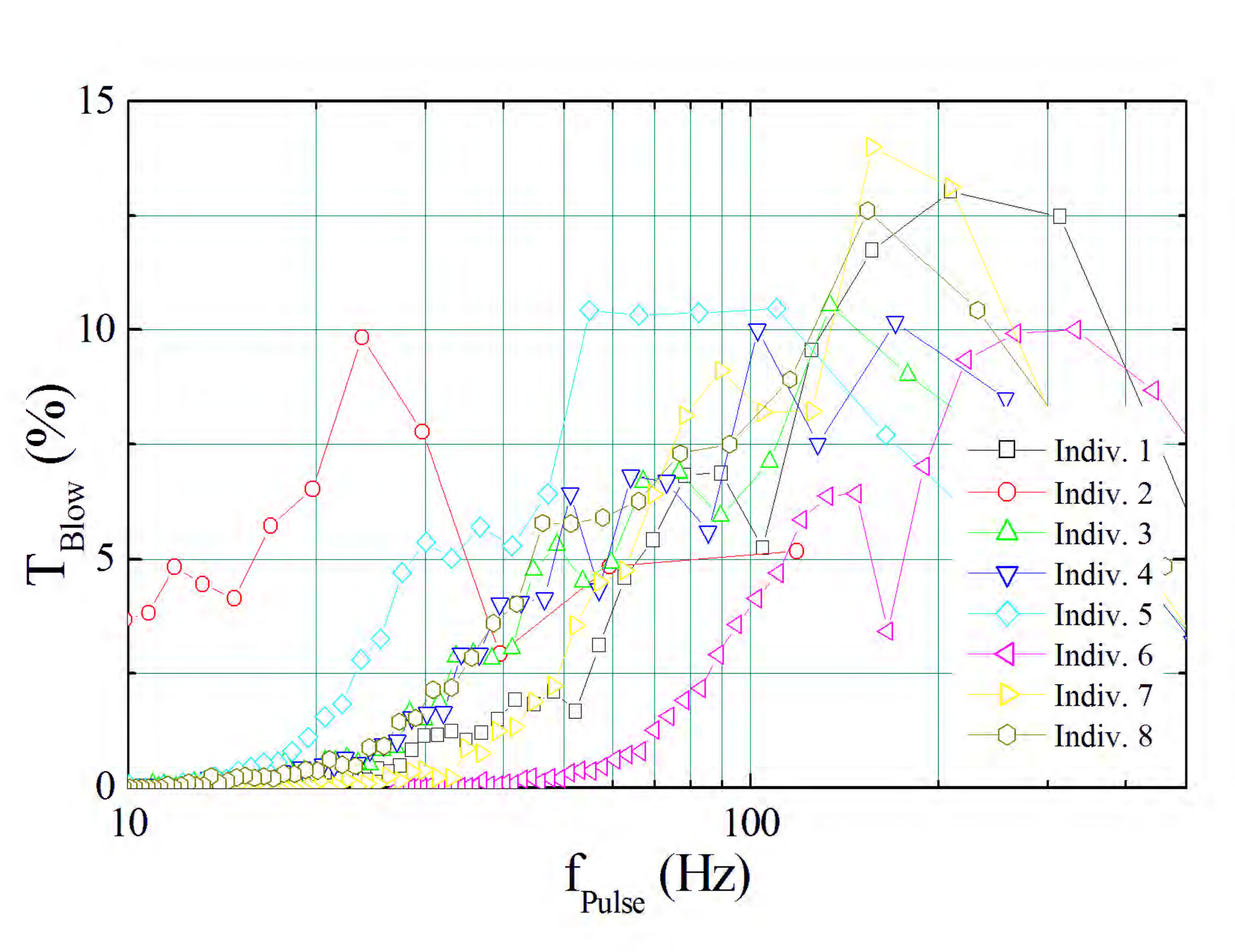}
\caption{Variation of the relative blowing times versus the observed pulse frequencies
 for the GPC individuals. Frequencies $f_{\mathrm{Pulse}}\geq100\,\text{Hz}$
 have to be treated with care due to sampling issues.
}
\label{fig:Tblow}
\end{figure}
To get a better understanding about the control laws synthesized by the 
GPC, the statistical distribution of the 
actuation frequency is analysed. For that purpose, 
the instantaneous actuation frequency is inferred via a zero-crossing 
algorithm. 
The actuation signal being updated at approximately 1\,kHz on the Arduino the discretization of the frequencies is larger at frequencies $f_{\mathrm{Pulse}}\geq 100$\,Hz.

Figure~\ref{fig:Tblow} shows the relative blowing time versus 
pulse frequency $f_{\mathrm{Pulse}}$ for all the GPC individuals. 
Except for individuals 2 and 5, the majority of the control laws 
show a major contribution of frequencies greater than 130\,Hz. 

Particularly, a major contribution of frequencies $f_{\mathrm{Pulse}}$ 
$\geq 250$\,Hz are observed for individual 6 which explains the 
origin of the low AVG blowing velocity due to the limited
switching speed of the valves. This leads to a significant amount
of time for which the valves are only partially opened.

For individual 2, the major frequency contribution is observed 
at $f_{\mathrm{Pulse}} \approx 24$\,Hz 
while individual $5$ presents a major frequency actuation in the 
range of $f_{\mathrm{Pulse}} =55 - 110$\,Hz. Those two individuals 
present a particular interest as previous experiments have shown that the Kelvin-Helmholtz instability is present close to separation point
at a frequency of $f_{\mathrm{KH}} = 110$\,Hz and a vortex shedding
at a frequency of about $f_{\mathrm{vs}} = 27 Hz$. Nevertheless, the
low relative $c_\mu$ induced during the individual 2 control process 
suggests that large periods without actuation occur, corroborated by
the bad cost function value of $J_{\mathrm P_{\mathrm{Stat}}}$.

\subsubsection{Separation length -- back flow area}
In this section the GPC actuation laws previously obtained are analysed in detail to
understand the mechanisms behind the best performing actuation laws.

The back flow area 
can be defined as the area in
which more than 50\% of the samples have a negative velocity. 
The resulting region is used to obtain the position of the attachment
and thus the separation length $L_{\mathrm{Sep}}$ (Tab.~\ref{tab:L_sep}) and
the size of the back flow area.
\begin{table}[!h]
\caption{Separation length and percentage of reduction with respect to the baseline for the best open-loop and the GPC individuals.}
\label{tab:L_sep}       % Give a unique label
\centering
\begin{tabular}{lll}
\hline\noalign{\smallskip}
& $L_{\mathrm{Sep}}/h$ & Reduction (\%)\\
\noalign{\smallskip}\hline\noalign{\smallskip}
Baseline & 5.4 & - \\
Best open-loop & 3.14 & 41.9 \\
Indiv. 1 & 3.31 & 38.7 \\
Indiv. 2 & 4.05 & 25.0 \\
Indiv. 3 & 3.43 & 36.5 \\
Indiv. 4 & 3.50 & 35.2 \\
Indiv. 5 & 3.12 & 42.2 \\
Indiv. 6 & 3.46 & 35.9 \\
Indiv. 7 & 3.16 & 41.5 \\
Indiv. 8 & 3.47 & 35.7 \\
\noalign{\smallskip}\hline
\end{tabular}
\end{table}
\begin{figure}[t]
  \includegraphics[width=0.45\textwidth]{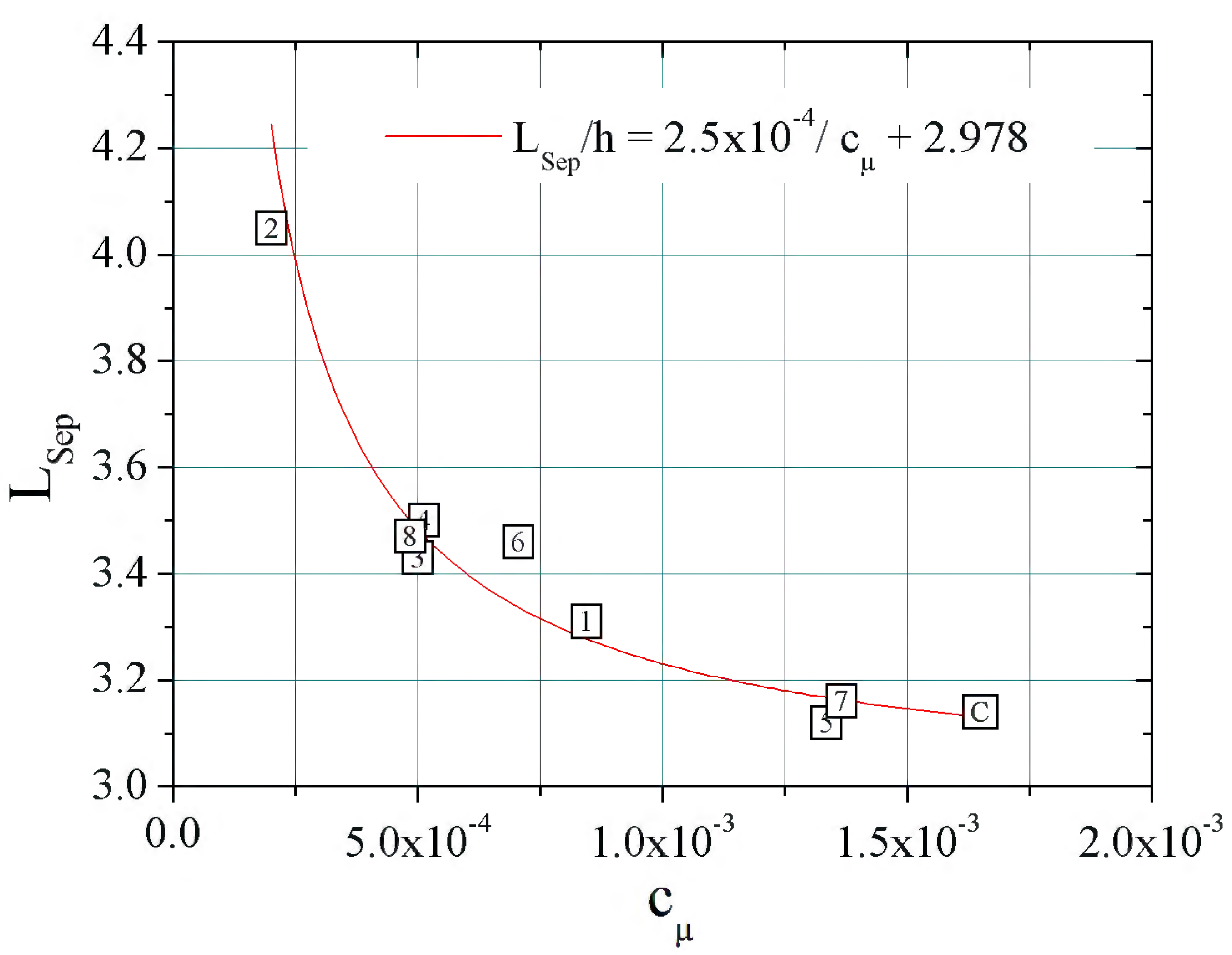}
\caption{Variation of the separation length versus the momentum coefficient for the best open-loop (label c) and the GPC individuals.
}
\label{fig:Lsep_cmu}
\end{figure}
\begin{figure}[h!]
  \includegraphics[width=0.45\textwidth]{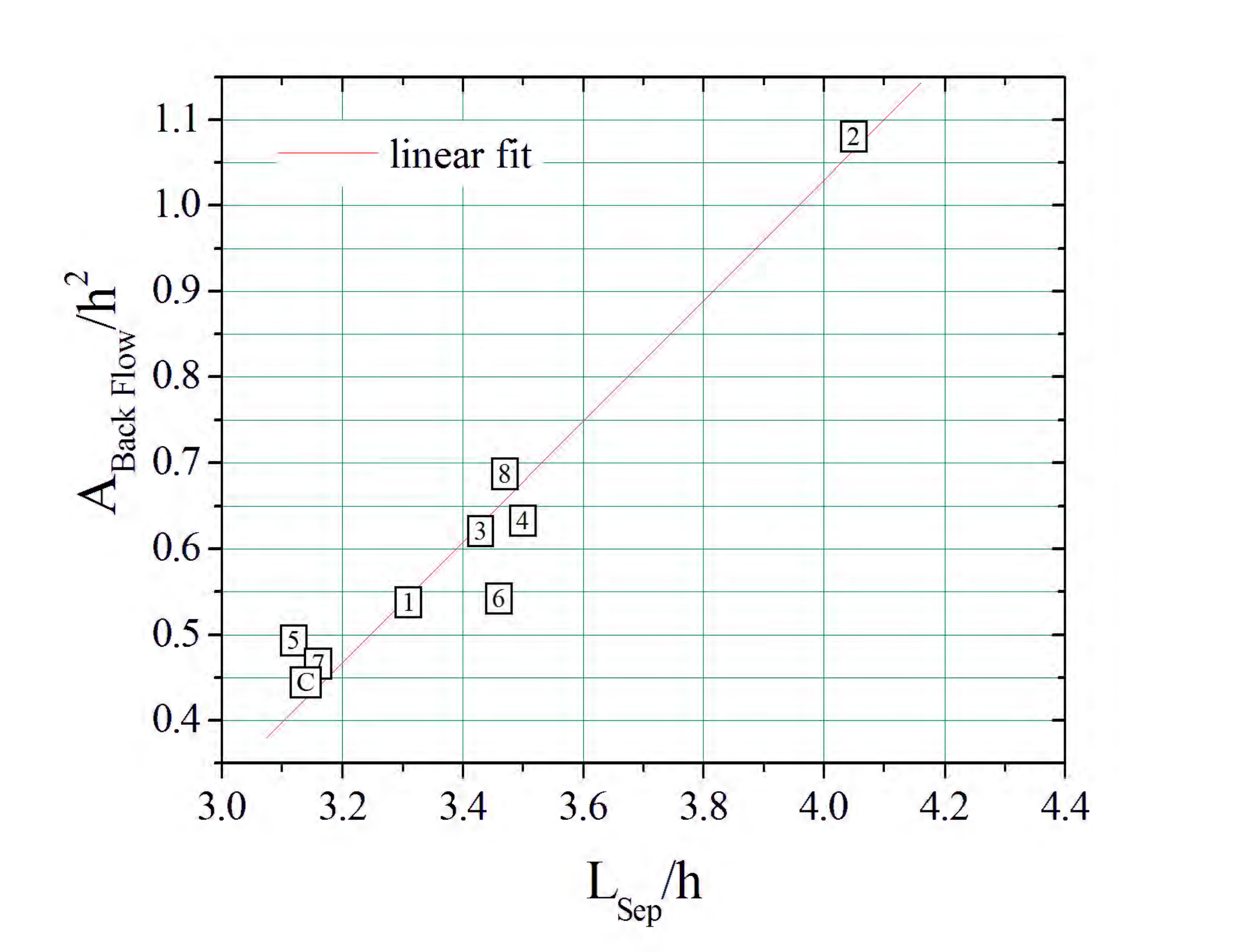}
\caption{Variation of the mean back flow area versus the separation length for the best open-loop (label c) and the GPC individuals.
}
\label{fig:Abf_Lsep}
\end{figure} 

When the best open-loop control is applied, the attachment point moves
upstream and the separation length decreases by 41.9\% compared to the
baseline case. For closed-loop control, $L_{\mathrm{Sep}}$ decreases by
25--42\% depending on the individual. Particularly, individuals 5 and
7 achieve almost the same separation length reduction (42 and 41.5\%,
respectively, see Tab.~\ref{tab:L_sep}) as the best open-loop solution while having
a lower $c_\mu$ consumption. Except for individual 6, the position of the
mean attachment point $L_{\mathrm{Sep}}/h$ against $c_\mu$
(Fig.~\ref{fig:Lsep_cmu}) presents an hyperbolic behaviour even though
the actuation laws and thus mechanisms differ from each other. 
\begin{figure}
  \includegraphics[width=0.45\textwidth]{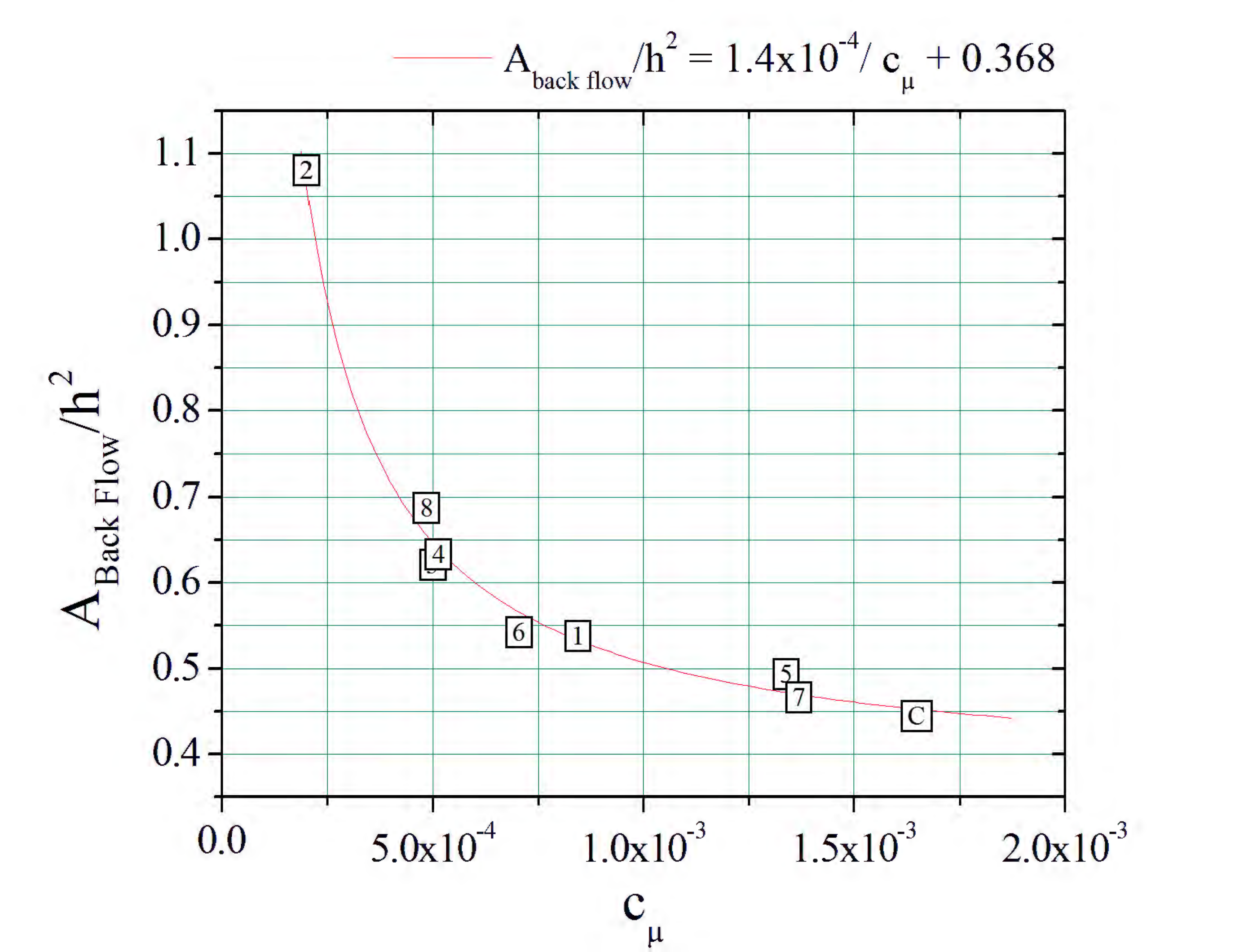}
\caption{Variation of the mean back flow area versus the momentum
  coefficient for the best open-loop (label c) and the GPC individuals.
}
\label{fig:Abf_cmu}
\end{figure}

In the same way, the back flow area is determined. Its evolution versus
the separation length (Fig.~\ref{fig:Abf_Lsep}) pre\-sents a linear
behaviour, except for individuals 5 and 6. This result shows that the
observation of the back flow area is equivalent to the observation of
the variation of the separation length. This is in agreement with the
recent work of \citet{Gautier2013}. Furthermore, the
monotonic evolution of the back flow area
versus the momentum coefficient (Fig.~\ref{fig:Abf_cmu}) shows
that in the given setup, the momentum coefficient seems to have the greatest
impact on the reduction of the separation length.

\subsubsection{Kinetic and turbulent kinetic energy flow rate}
In this section, we discuss the GPC individuals in terms of their energy content.
For that purpose, we introduce the mean kinetic energy flow rate ($M_{E_c}$) 
and the turbulent kinetic energy flow rate ($M_k$) respectively defined as:
\begin{align}
M_{E_c} &= \int_{S} (\vec{U}(x)\cdot\vec{n})\,E_c(x)\,\mathrm{d}S(x), \quad\text{and}
  \label{eq:M_Ec}\\
M_{k} &= \int_{S} (\vec{U}(x)\cdot\vec{n})\,k(x)\,\mathrm{d}S(x).
  \label{eq:M_k}
\end{align}
where $\vec{U}$ is the mean velocity and $\vec{n}$ denotes the outward-pointing normal to $S$. $E_c$ represents the 
energy transported by the mean velocity and $k$ is the turbulent kinetic energy (TKE). 
Since the turbulent production term is expected to be damped when the recirculation region is reduced, the energy transported by the mean velocity 
downstream of the reattachment point is expected to be larger when control occurs.
If one applies flow control, $M_k$ is governed by two properties: the
modification of TKE value in the wake and the thickness of the wake. Thus,
a control with a small $c_\mu$ should increase the TKE by
increasing the mixing without reducing the wake thickness. As $c_\mu$
increases further, the thickness of the wake should decrease, leading to
a decrease of $M_k$.
\begin{figure}[!b]
  \includegraphics[width=0.45\textwidth]{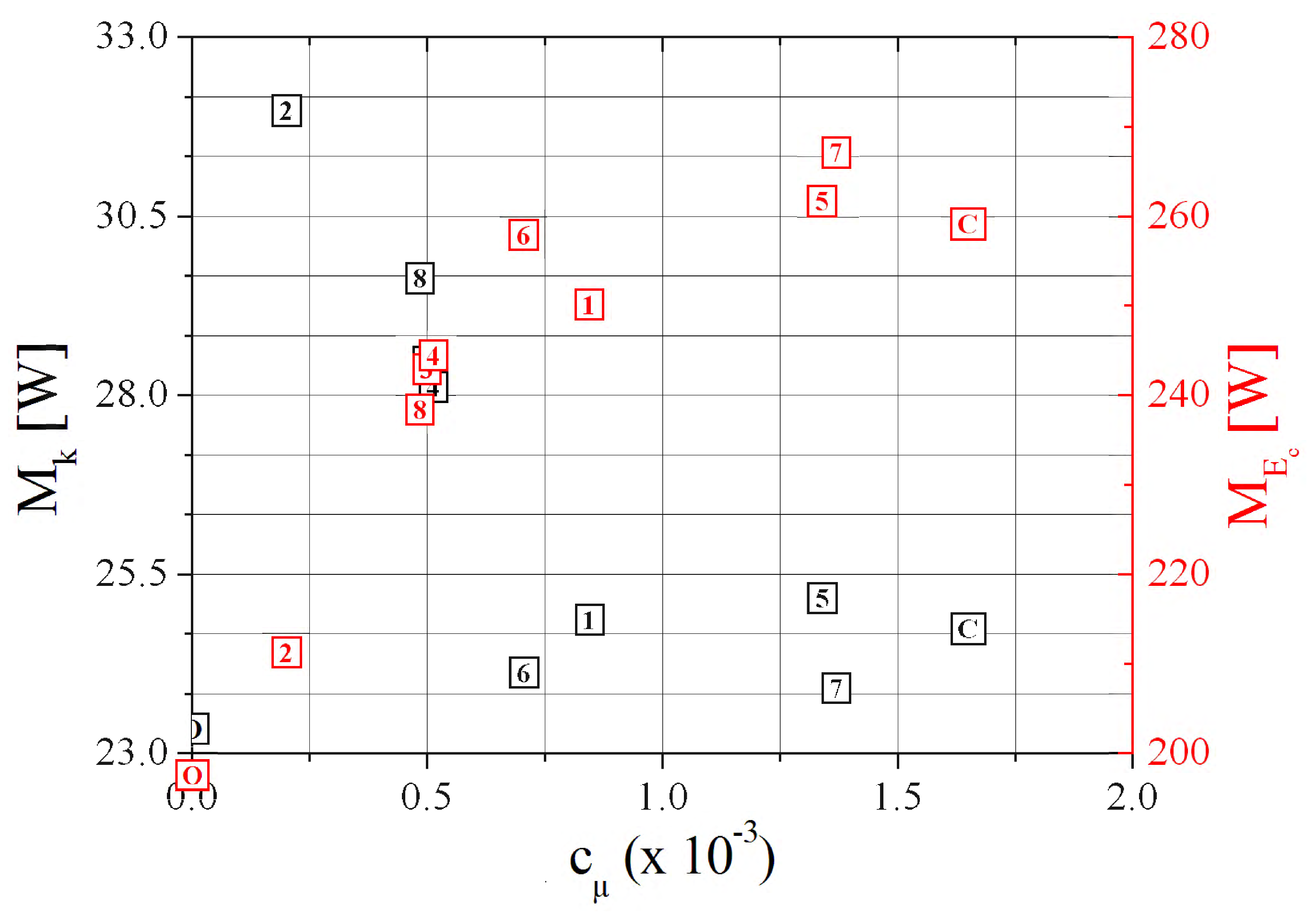}
\caption{Variation of the turbulent kinetic energy flow rate $M_k$ and mean kinetic energy flow rate $M_{E_c}$ versus the momentum coefficient at $x/h = 5.5$ for the baseline (label o), for the best open-loop (label c) and for the GPC individuals.
}
\label{fig:Mk_cmu}
\end{figure}

\begin{figure}[!t]
  \includegraphics[width=0.45\textwidth]{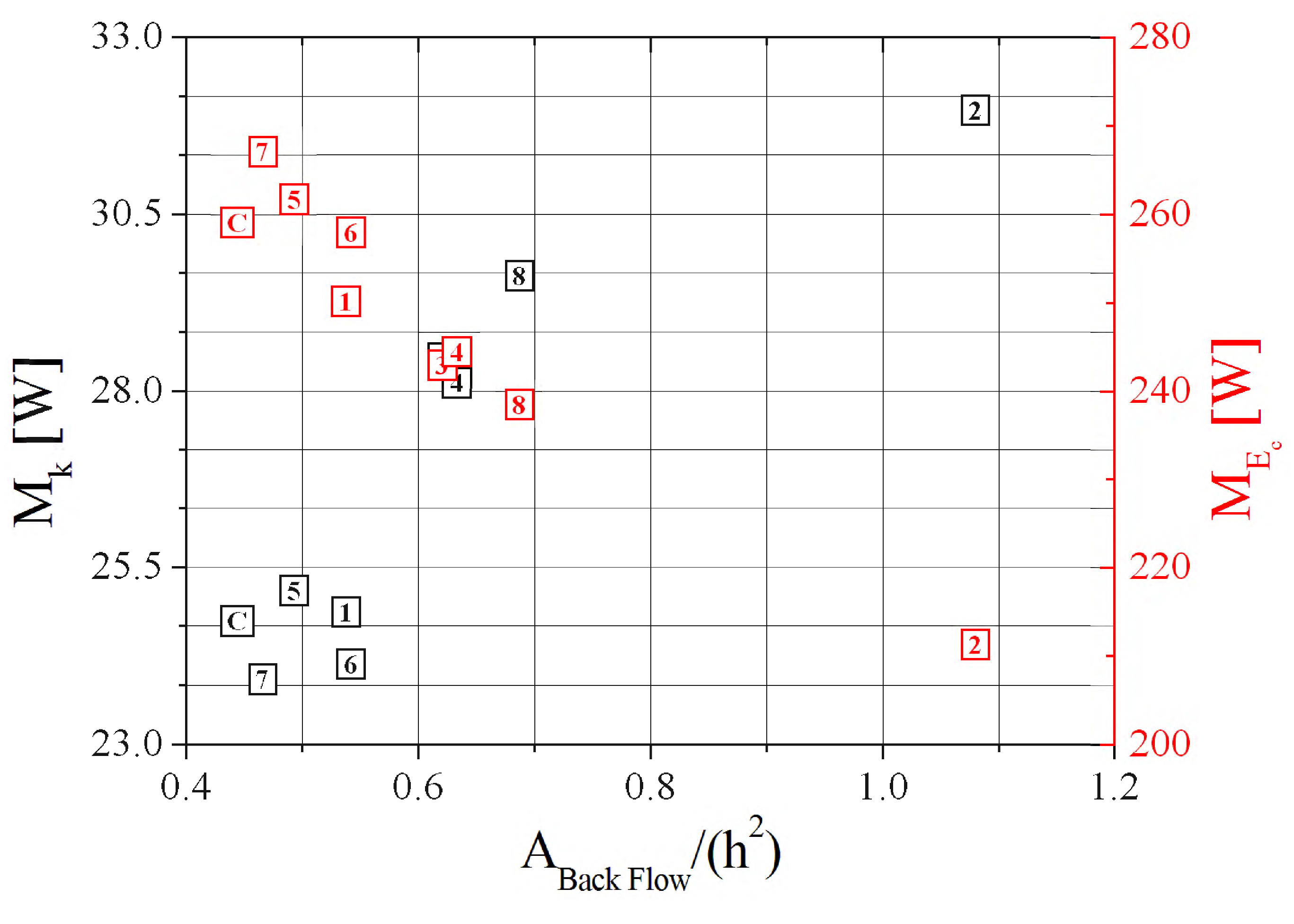}
\caption{Variation of the turbulent kinetic energy flow rate $M_k$ and mean kinetic energy flow rate $M_{E_c}$ versus the mean back flow area at $x/h = 5.5$ for the best open-loop (label c) and for the GPC individuals.
}
\label{fig:Mk_Abf}
\end{figure}

Figures~\ref{fig:Mk_cmu} and~\ref{fig:Mk_Abf} show $M_{E_c}$ and
$M_k$ (at $x/h$ = $5.5$) versus $c_\mu$ and respectively versus
the size of back flow area. 
$M_{E_c}$ versus $c_\mu$ does not exhibit a clear trend (Fig.~\ref
{fig:Mk_cmu}). First a large increase is observed for $c_\mu \leq 
8.45 \times 10^{-4}$. For $c_\mu > 8.45 \times 10^{-4}$ the data
levels out in a plateau with 3\% fluctations in $M_{E_c}$. In 
contrast, $M_{E_c}$ decreases linearly against the back flow area as
seen in Fig.~\ref{fig:Mk_Abf}. Furthermore, for similar back flow area
value, this curve highlights more efficient individuals with a large
$M_{E_c}$, for example individual $7$ compared to the best 
open-loop control or individual $6$ compared to individual $1$.

This efficiency variation could be related to two parameters:
the shaping of the wake caused by the form of a
particular control law or the overall mass flow value $c_\mu$
used by the control law.
For individual 7 and the best open-loop, the control process could suggest 
that the physical process involved to control the flow could be responsible 
for $M_{E_c}$'s efficiency variation. Nevertheless, Fig.~\ref{fig:Tblow} 
shows that frequencies injected in the flow for individuals 1 and 6 
present a similar broad bandwidth but $c_\mu$ is decreased 
by 17\% for individual 6. As mentioned earlier, this large 
reduction of $c_\mu$ is due to low $V_\mathrm{Jet}$ induced by the 
time response of the valves. For this case, a 17\% $c_\mu$ 
reduction is responsible for a 3\% $M_{E_c}$ increase.

The evolution of $M_k$ versus $c_\mu$ is related to $M_{E_c}$: for
$c_\mu \leq 8.45 \times 10^{-4}$, $M_k$ decreases quickly with growing
$c_\mu$ and for $c_\mu>8.45\times 10^{-4}$ it levels off in a plateau-like
shape with 5\% fluctuations. The evolution of $M_k$ versus the back flow
area present a global trend where $M_k$ increases with the back flow
area, as expected. Furthermore, the evolution of $M_{E_c}$ and 
$M_k$ versus $c_\mu$ indicates the existence of a plateau beginning at a $c_\mu$ value greater than the $c_\mu$ value of individual 6 ($7.10^{-4}$), confirming its efficiency regarding those parameters.

Analogous to $M_\mathrm{Jet}$, a criterium $M_k$ given by \eqref{eq:M_k} is introduced 
using the TKE flow rate, which allows to obtain a coefficient $C_{\mathrm{Jet},k}$ given by:
\begin{equation}
C_{\mathrm{Jet},k} = \frac{M_\mathrm{Jet}}{M_k}.  
\label{eq:C_jk}
\end{equation}

\begin{figure}
  \includegraphics[width=0.45\textwidth]{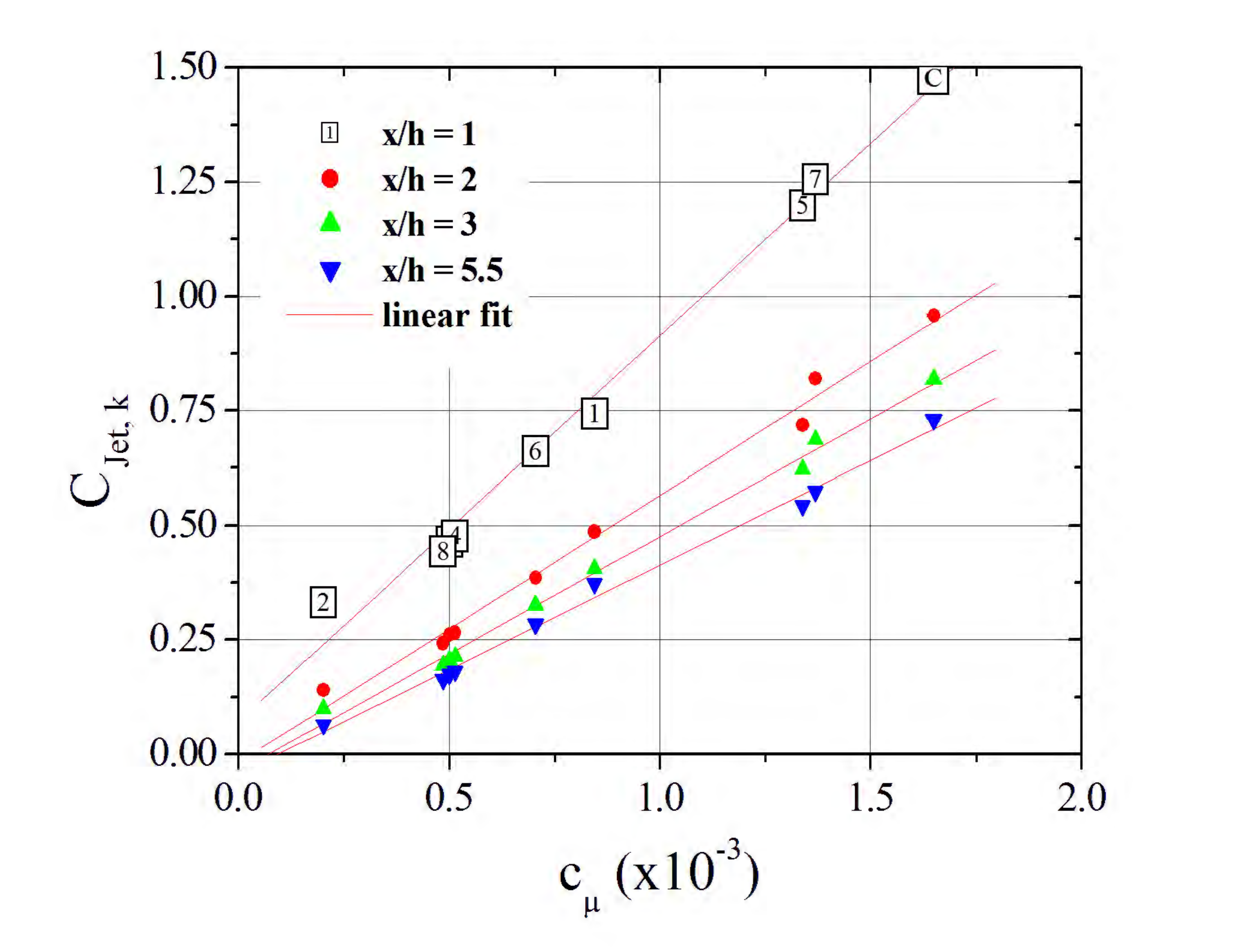}
\caption{Variation of the coefficient $C_{\text{Jet},k}$ versus the momentum coefficient at four streamwise locations for the best open-loop (label c) and for the GPC individuals. 
}
\label{fig:Cjk_Cmu}
\end{figure}

This coefficient $C_{\mathrm{Jet},k}$ is plotted versus $c_\mu$ 
in Fig.~\ref{fig:Cjk_Cmu}. It appears that the relation between $C_{\mathrm{Jet},k}$ and 
$c_\mu$ is linear. To explain this relation, we can first refer to the relation 
of $M_\mathrm{Jet}$ and $M_k$ versus $c_\mu$. Considering a quasi constant $V_\mathrm{Jet}$ blowing, 
$M_\mathrm{Jet} \propto c_\mu$ while $M_k \propto c_\mu ^{-1}$ for $c_\mu \leq 8.45 \times 
10^{-4}$ and $M_k \propto  c_\mu ^{0}$ for $c_\mu \geq 8.45 \times 10^{-4}$. 
Furthermore, comparing the relative variation $\Delta M_\mathrm{Jet}$ ( $\sim \frac{ 
M_{\mathrm{Jet}_{\mathrm{Max}}} - M_{\mathrm{Jet}_{\mathrm{Min}}} }{\overline{M_\mathrm{Jet}}}$ ) and $\Delta 
M_k$, it appears that $\Delta M_\mathrm{Jet} \approx 10  \Delta M_k$. So $C_{\mathrm{Jet},k}$ is 
governed by $M_\mathrm{Jet}$ which presents a linear relation with $c_\mu$.

The study of the different individuals shows that GPC is able
to provide three individuals which achieved the best performance with respect to their respective cost-function. Also, it shows that accordingly to the goal to be achieved GPC explores the different mechanisms that can be used to optimize the cost function. If one looks at the separation properties, individuals 5 and 7
appear to be the most efficient individuals whereas if one looks at the energy
flow rate, individual 6 showed the best performance. It shows a reduction in
$c_\mu$ of 46\% compared to open-loop whereas individuals 5 and 7 have a similar
$c_\mu$ to open-loop. Consequently, the separation length of individual 6 is
larger than the open-loop case.
\begin{figure*}[!t]
\includegraphics[width=0.8\textwidth]{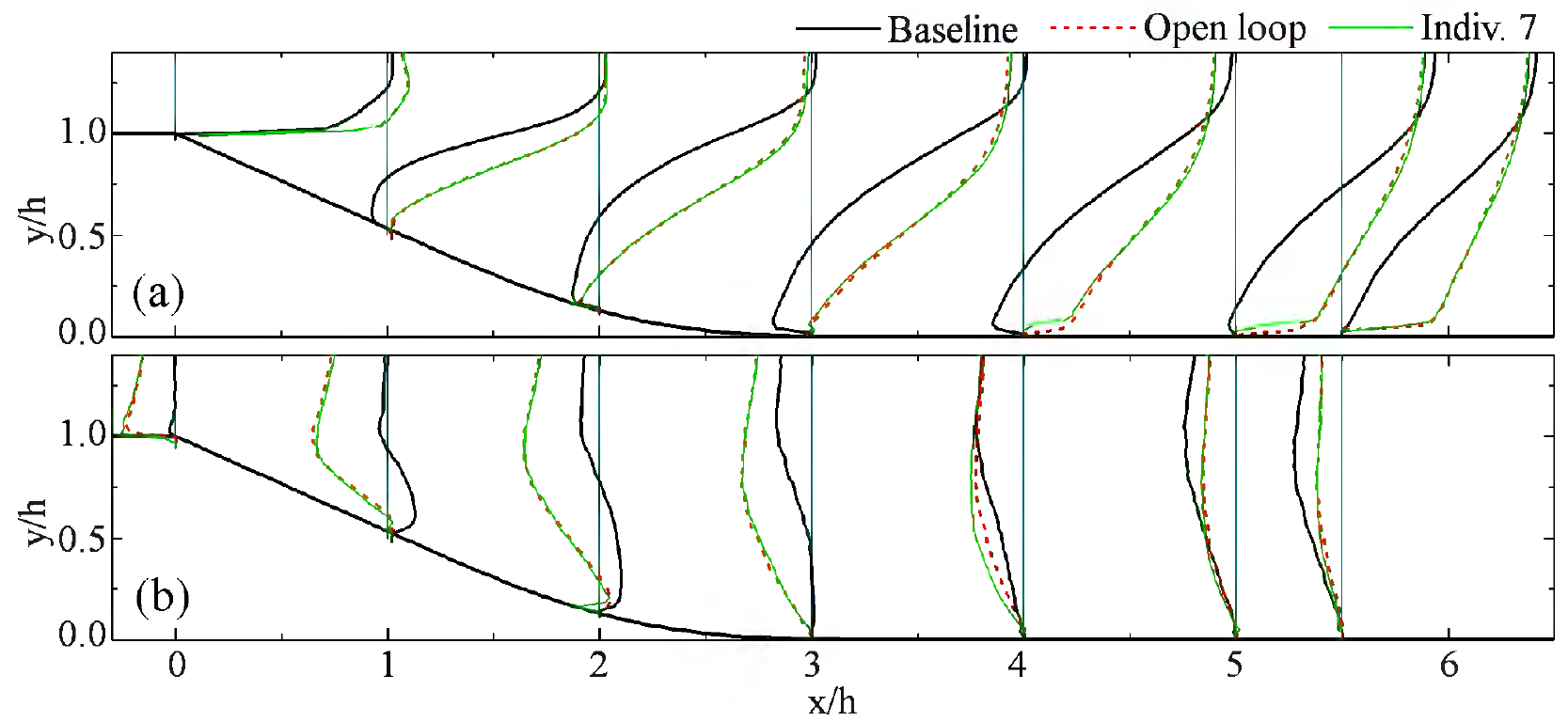}
\caption{Mean streamwise velocity $x/h + U/U_{\infty}$ (a) and wall-normal velocity $x/h + 2.5 \times V/U_{\infty}$ (b) over the sharp edge ramp for the baseline, best open-loop and GPC individual 7.} 
\label{fig:PlotAvgPrf}       % Give a unique label
\end{figure*}

\subsection{Analysis of the individual 7}
\label{sec:controlled_flow_analysis}

As observed above, the individual 7 can be ranked among the most efficient solutions 
and presents a similar $c_\mu$ value to the best open-loop control. For these reasons, an 
in-depth flow analysis of the individual 7 is now performed by comparison to the best open-loop control.

The expression for this individual is given below:
\begin{align}
%-1.23\frac{0.59 + \langle\mathrm{HF}_1\rangle + \mathrm{HF}_1 + 
%\tanh(\mathrm{HF}_1)}{\mathrm{HF}_1\mathrm{HF}_2}  
%  + \ln\left(\frac{\mathrm{HF}_2 + \mathrm{HF}_1 - \langle\mathrm{HF}%_1\rangle}{\langle\mathrm{HF}_2\rangle + \mathrm{HF}_2} \right).
b(t)=-1.23\frac{0.59 + s_1 + s_2 + \tanh(s_2)}{s_2s_5}  
 + \ln\left(\frac{s_5 + s_3}{s_4 + s_5} \right).
\end{align}
The division and the natural logarithm are capped so that there exist
no division by 0 as explained in Sec.~\ref{sec:MLC}. 

\begin{figure}[!t]
  \includegraphics[width=0.45\textwidth]{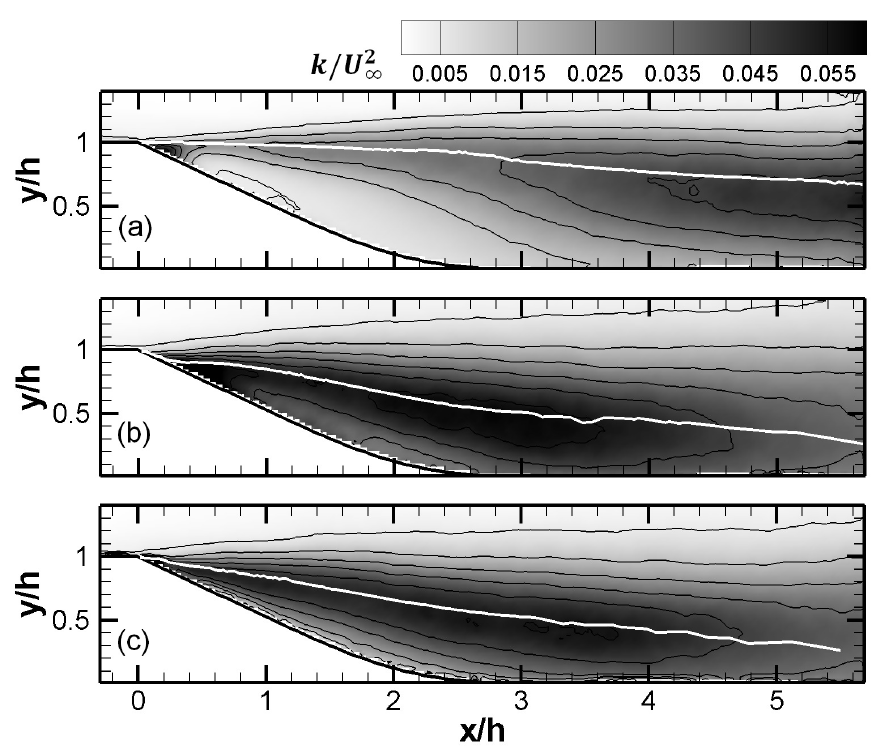}
\caption{Turbulent kinetic energy for the baseline (a), the best open-loop (b) and the GPC individual 7 (c). White line corresponds to the center line of the shear layer.
}
\label{fig:VFtke}
\end{figure}

As shown in Fig.~\ref{fig:PlotAvgPrf}(a), AVGs promote the momentum transfer between the 
near wall and the free stream yielding a larger mean streamwise velocity in the 
neighborhood of the sharp edge. 
In addition, the counter-rotating vortices generated by the AVGs induce a significant sweep 
flow motion at the laser sheet position. This is evidenced by the strong negative mean 
transverse velocity close to the sharp edge. Combined together, these mechanisms delay the 
separation and deflect the separated shear layer towards the wall yielding a shorter 
recirculation (see Fig.~\ref{fig:VFtke}).

For the baseline case, a higher level of TKE 
(see Fig.~\ref{fig:VFtke} (a)) is observed close to the center line of the shear
layer which is progressively deflected towards the wall.
At the attachment point, the angle of deflection of shear layer corresponds to
3.3\textdegree.
The maximum TKE value $k/U_{\infty}^2 \approx 0.56$ is achieved close to the
separation point. It decreases quickly down to a minimal value of
$k/U_{\infty} ^2 \approx 0.24$ in the shear layer at $x/h \approx 0.63$. Beyond
this point, the TKE increases in the shear layer up to the reattachment 
point ($k/U_{\infty} ^2 \approx 0.5$).
On the other hand, the recirculation bubble below the shear layer is characterized
by a low level of TKE ($5 \times 10^{-3} \leq k/U_{\infty}^2 \leq 1 \times 10^{-2}$).

For the best open-loop, the evolution of the TKE in the shear layer is similar
to the baseline case but higher levels are observed downstream from the 
separation point ($k/U_{\infty}^2 \approx 0.75$).
This maximal value decreases up to $x/h =1$ ($k/U_{\infty} ^2 \approx 0.54$) and
then increases progressively up to the attachment point 
($k/U_{\infty} ^2 \approx 0.58$). At the attachment point, the shear layer
center line shows a larger angle of deflection of 9.6\textdegree. This 
intensification of the TKE shows also in the recirculation bubble where 
$0.3 \leq k/U_{\infty} ^2 \leq 0.5$.
As already observed in \citet{Debien2015TSFP9}, this intensification in the TKE is due to the lock-on 
of vortex shedding by the actuators, which is responsible for the generation of
large spanwise structures along the ramp.

For the individual 7, the evolution of the TKE in
the shear layer presents an important difference since a progressive increase
from separation point ($k/U_{\infty} ^2 \approx 0.5$) up to reattachment point 
($k/U_{\infty} ^2 \approx 0.54$) occurs. At the reattachment point, the shear layer center line
shows also a large angle of deflection of 9\textdegree. The recirculation bubble
presents also an intensification in the TKE level but a thin layer with low TKE level 
($0.1 \leq k/U_{\infty}^2 \leq 0.2$) is present above the ramp from $x/h \geq 0.55$
up to attachment point.
It appears that contrary to the best open-loop control where large scale structures
are generated by the AVGs blowing to enhance mixing in recirculation bubble, 
the GPC control strategy seems to stimulate the development of shear layer and
to stabilize it as it does not show maximal TKE level near separation point but
a fast growth of TKE level is achieved downstream of the separation point.

%\subsubsection{shear layer and separation bubble analysis}
\begin{figure}[!h]
  \includegraphics[width=0.45\textwidth]{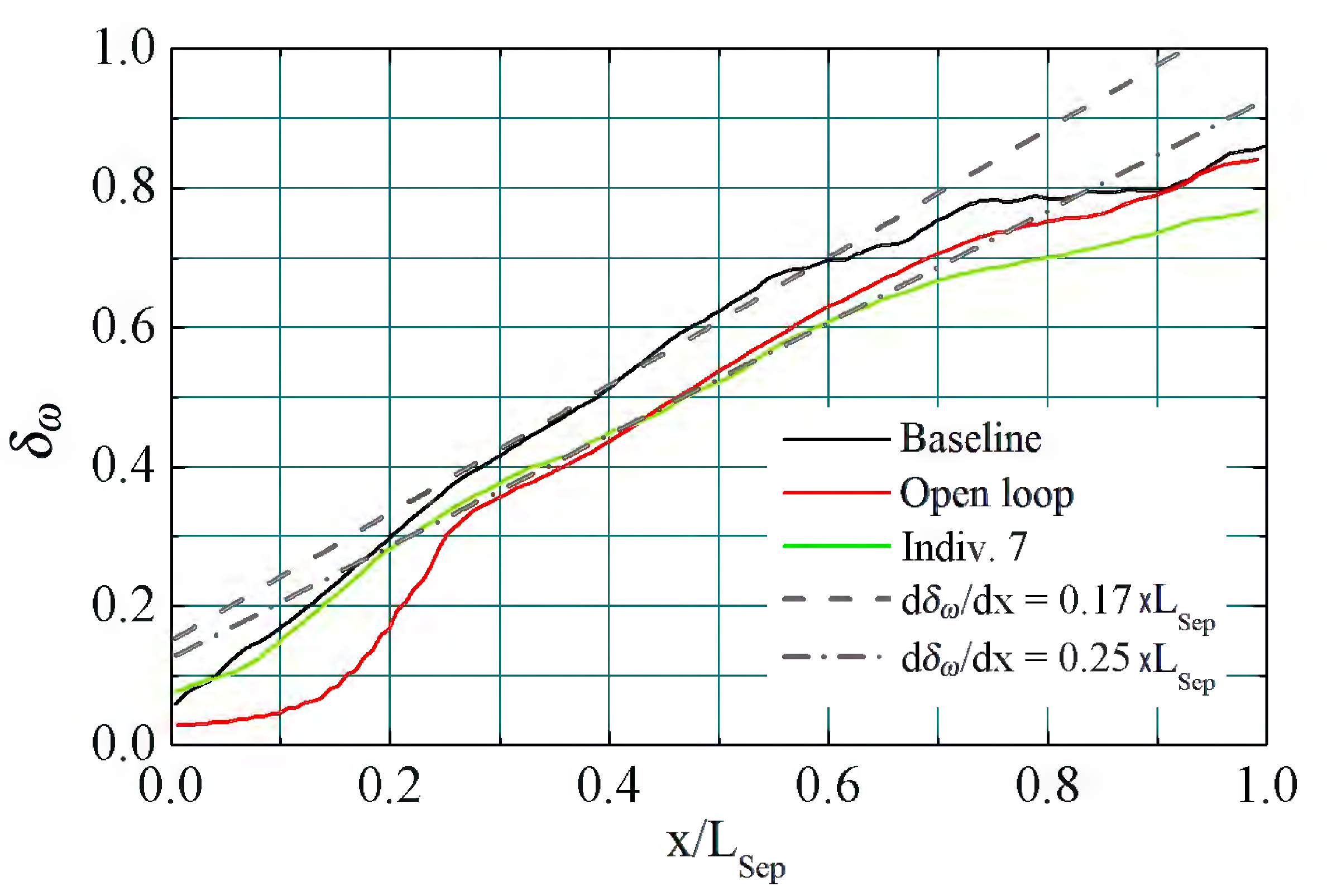}
\caption{Streamwise variation of the vorticity thickness over the ramp for baseline, best open-loop and GPC individual 7.
}
\label{fig:DeltaW}
\end{figure}

\begin{figure*}
\begin{center}
\includegraphics[width=0.7\textwidth]{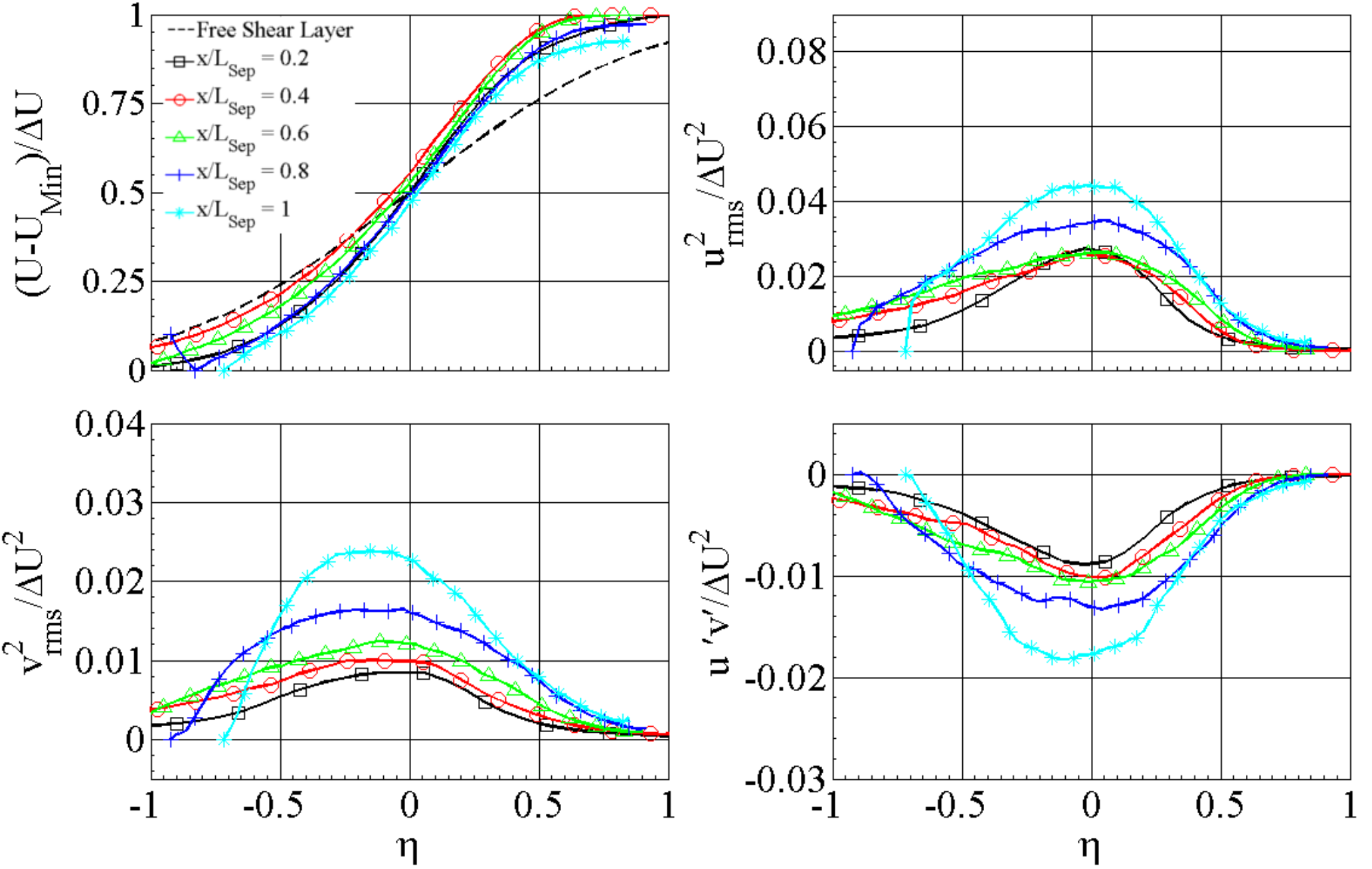}
\end{center}
\caption{Mean streamwise velocity and Reynolds stresses profiles of the
  mixing layer in similarity coordinates for the baseline.
}
\label{fig:ShLyBsl}
\end{figure*}
\begin{figure*}
\begin{center}
\includegraphics[width=0.7\textwidth]{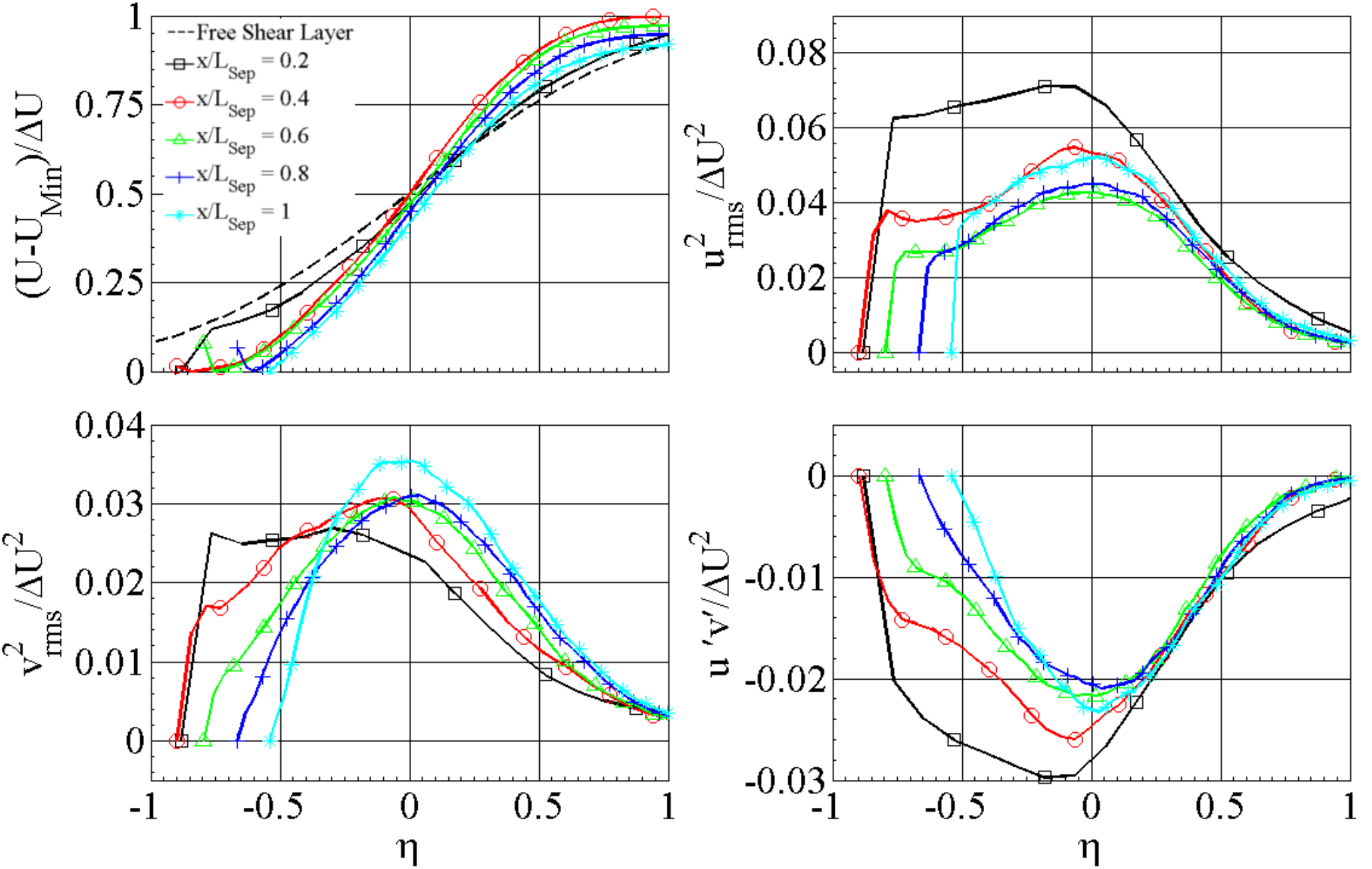}
\end{center}
\caption{Mean streamwise velocity and Reynolds stresses profiles of the
  mixing layer in similarity coordinates for the best open-loop.
}
\label{fig:ShLyC30}
\end{figure*}
\begin{figure*}
\begin{center}
\includegraphics[width=0.7\textwidth]{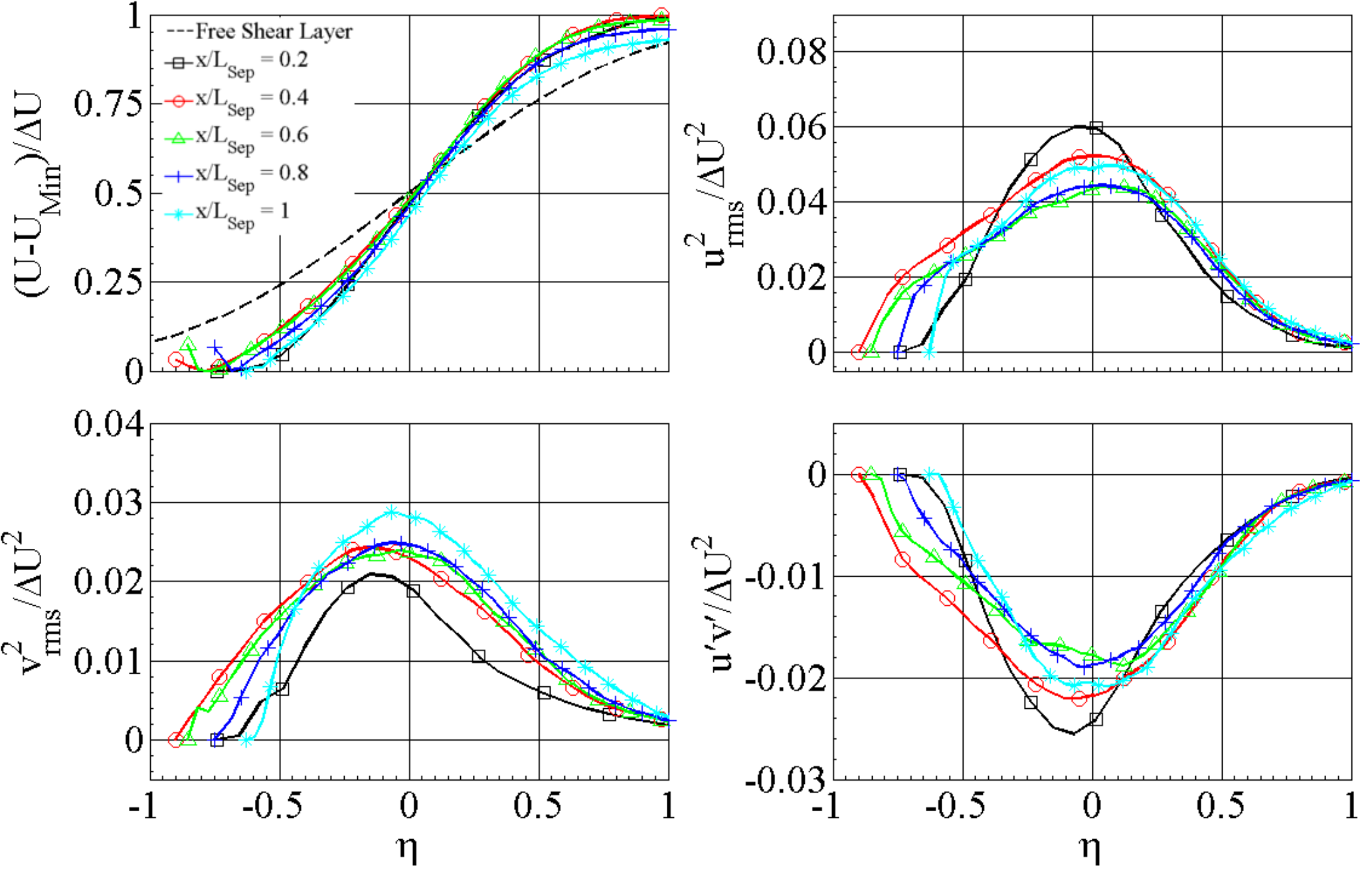}
\end{center}
\caption{Mean streamwise velocity and Reynolds stresses profiles of the
  mixing layer in similarity coordinates for the GPC individual 7.
}
\label{fig:ShLyIndiv7}
\end{figure*}

The evolution of the vorticity thickness ($\delta_{\omega}$) of the shear layer 
above the separation zone is presented Fig.~\ref{fig:DeltaW} for the baseline, 
the best open-loop and the GPC individual. As observed by \citet{Castro1987} and 
\citet{Jovic1996}, the vorticity thickness 
is featured by a large growth rate which progressively decreases up to the 
reattachment point. Using a linear fit in the least-square sense, the growth rate of the 
separated shear-layer for both the baseline and the best open-loop solution cases is
$\tfrac{\partial \delta_{\omega}}{\partial (x/h)} = 0.17$ over the range $0.25 \leq x/L_{\mathrm{Sep}} \leq 0.6$. This value is close to that reported for free shear-layers (see e.g.\citep{Browand1985}). The growth rate of the separated shear-layer induced by the individual 7 seems to be larger $\tfrac{\partial 
\delta_{\omega}}{\partial (x/h)} = 0.25$ (see Fig.~\ref{fig:DeltaW}). 
Note that unlike baseline and the GPC individuals, the best open-loop control 
vorticity thickness grows exponentially up to $x/L_{\mathrm{Sep}} = 
0.25$ before its linear growth. This change coincides with the occurrence of high levels of TKE. For the best open-loop control, a low actuation frequency is used to
lock on the shedding frequency. Streamwise vortices are supposed to be produced
by the low AVGs' activation frequency, and these vortices are able to develop
up to $x/L_{\mathrm{Sep}} \approx 0.25$ despite 
the presence of the sharp edge \citep{Debien2015TSFP9}. Beyond $x/L_{\mathrm{Sep}} \geq 0.25$, the classic development of the shear layer occurs.

In order to compare the shear-layers developing above the
recirculation region, mean velocity profiles and Reynolds stresses are
plotted in dimensionless form in Figs.~\ref{fig:ShLyBsl}, \ref{fig:ShLyC30} and \ref{fig:ShLyIndiv7} for the baseline, the open-loop control and the closed-loop control cases, respectively. For that purpose, the reduced coordinate $\eta = \tfrac{y-y_c}{\delta_{\omega}}$ is 
used, where $y_c$ corresponds to the position of maximal velocity gradient (see 
Fig.~\ref{fig:VFtke} in which the white line corresponds to $y_c$). The velocity 
difference $\Delta U=U_{\infty}-U_{\mathrm{Min}}$ is used as the reference 
velocity, where $U_{\mathrm{Min}}$ is the minimum streamwise velocity of the profile. 
For the baseline case (Fig.~\ref{fig:ShLyBsl}), the dimensionless 
mean streamwise velocity profiles do not reach a self-similar state which seems to be
induced by the presence of the separation bubble which mainly affects the evolution of the 
lower part of velocity profiles ($\eta \leq 0 $) and of the maximal velocity 
gradient along the ramp. For the best open-loop (Fig.~\ref{fig:ShLyC30}) and 
the individual 7 (Fig.~\ref{fig:ShLyIndiv7}), the
dimensionless mean streamwise velocity profiles collapse downstream from
$x/L_{\mathrm{Sep}} \geq 0.25$, achieving a nearly self-similar state. 

The baseline case presents higher Reynolds stresses levels close to the 
shear layer center line \citep{Castro1987} while low Reynolds 
stresses levels are observed close to the wall in the separation bubble, 
consistent with the literature \citep{Song2000IJHFF}. From the sharp edge ramp, the 
high level of $u^2_{\mathrm{rms}}/\Delta U^2$ in the shear layer progressively 
decreases up to $0.024$ at $x/L_{\mathrm{Sep}} = 
0.5$ and then increases up to reattachment point. In contrast, 
$v^2_{\mathrm{rms}}/\Delta U^2$ and $u'v'/\Delta U^2$ increase from separation 
point up to attachment point where $v^2_{\mathrm{rms}}/\Delta U^2$ $= 0.028$ and 
$u'v'/\Delta U^2 = -0.018$.

The actuation introduced by the AVGs increases the Reynolds stresses levels in the recirculating 
bubble. In the shear layer, both open-loop and indiv.\ 7 have similar characteristics. 
$u^2_{\mathrm{rms}}/\Delta U^2$ presents higher level close to the separation 
point, decreases up to $x/L_{\mathrm{Sep}} = 0.6$ and then increases up to the 
attachment point. $v^2_{\mathrm{rms}}/\Delta U^2$ presents an increase from 
separation point up to attachment point. Downstream a plateau at $ 0.4 \leq 
x/L_{\mathrm{Sep}} \leq 0.8$ for open-loop, and $ 0.4 \leq x/L_{\mathrm{Sep}} 
\leq 0.5$ for individual $7$ is visible. The shear component $u'v'/\Delta U^2$
decreases up to $x/L_{\mathrm{Sep}} = 0.7$ for both cases. Furthermore, the local
maximum of $\frac{v^2_{\mathrm{rms}}}{\Delta U^2}$ and $\frac{u'v'}{\Delta U^2}$ are 
shifted towards the lower part of the shear layer ($\eta \leq 0$) close to the 
separation point (see $x/L_{\mathrm{Sep}} = 0.2$, Fig.~\ref{fig:ShLyC30} 
and~\ref{fig:ShLyIndiv7}). For the best open-loop control, near wall Reynolds stresses  
present high value for $x/L_{\mathrm{Sep}} \leq 0.6$ due to the convection of 
large vortices created by the AVGs \citep{Debien2015TSFP9}. 

The dimensionless Reynolds stresses reveal drastic changes 
between the baseline and the controlled cases. The decrease of the shear Reynolds stresses 
component up to $x/L_{\mathrm{Sep}} = 0.7$ when control is applied suggests that
the shear layer development is stimulated by AVGs' blowing. This could explain
the self-similar state of the dimensionless streamwise mean velocity achieved with
flow control implying that the driving of the shear layer development is due to
the convection and growth of structures induced by the AVGs' blowing.

\begin{figure}
 \includegraphics[width=0.45\textwidth]{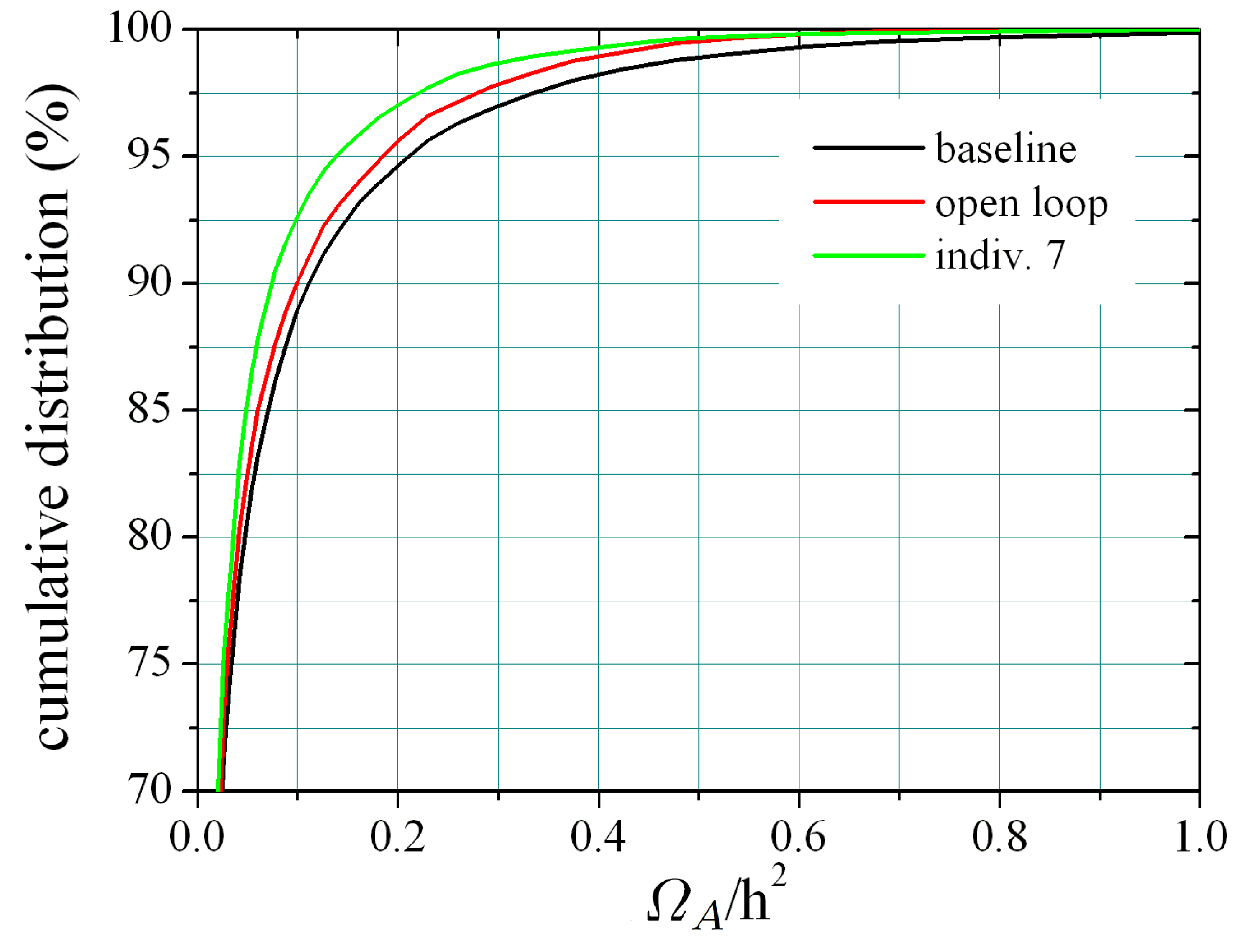}
\caption{Cumulative distribution of area of the detected vortex regions for the baseline, the best open-loop and GPC individual 7.
}
\label{fig:Vtx_size}
\end{figure}

The extraction of the vortex region $\Omega$ is now obtained using the vector field 
topology. The area of the detected vortex region $\Omega_A$ is extracted to obtain the 
vortex area distribution that is presented in Fig.~\ref{fig:Vtx_size} for the baseline 
case, the best open-loop control and the individual 7. Overall, 50\% of the detected 
vortices occupy an effective area $\Omega_A$ smaller than $0.01 h^2$.
The cumulative distributions plotted in Fig.~\ref{fig:Vtx_size} evidence that the GPC 
control induce the production of smaller vortices compared to both baseline and open-loop 
control. This property may somehow be related to the broad range of frequencies excited by 
the GPC control law. For the open-loop case, we have reported elsewhere (see 
\citet{Debien2015TSFP9}) a lock-on mechanism responsible for the generation of shedded 
large-scale structures. Therefore, even though both control strategies mainly act on the 
growth of the separated shear-layer, the results displayed in Fig.~\ref{fig:Vtx_size} 
suggest two different mechanisms leading to a nearly identical mean separation region: i. a 
rapid growth of the shear-layer due to large-scale structure engulfment (open-loop) and ii. 
an enhancement of the local mixing of the shear-layer (GPC).

\section{Conclusions}
\label{sec:conclusions}
Genetic programming control (GPC) has been applied to closed-loop forcing
of a separated turbulent boundary layer over a sharp edge ramp 
using only the signal of two hot-film sensors placed near the inflection point of the ramp and static pressure sensors along the ramp. 
GPC achieves control laws aiming to reduce the separation length 
with a penalized momentum  coefficient.
The resulting law minimizes a given cost function.
The performance is monitored by the pressure distribution, 
the hot-film signals and the momentum coefficient.  
By varying the actuation penalization of the cost function,
multiple optimization points were obtained.

The performance of GPC is benchmarked
with the optimized periodic actuation.
The reduction of the separation length, the back flow area, 
and the pressure distribution is similar 
but GPC achieves this separation mitigation
with a smaller momentum coefficient. 
This can be explained by the direct relation 
between the momentum coefficient, the separation length and the back flow area. 
Furthermore, the kinetic energy of the mean flow field reveals that GPC achieves a 
better increase in the kinetic energy than the best open-loop control. 

The velocity field properties were analyzed using a Stereo PIV system.
Similar performance benefits are obtained with open and closed-loop control. 
For the best open-loop control, 
the actuation frequency is chosen close to shedding mode to obtain a lock-on control. 
Downstream, the flow displays 
the streamwise vortices signature induced by the AVGs 
up to $x/L_{\mathrm{Sep}} = 0.25$. 
Beyond this point, the classic shear layer develops 
 and presents a similarity state for the mean streamwise velocity, 
induced by the growth and convection of large vortex region 
achieving the reduction of separation length.
In contrast, 
the analysis of the best GPC laws show 
much higher actuation frequencies from twice to ten times the best open-loop frequency. 
The vector field analysis also reveals 
that streamwise vorticity signature induced by the AVGs 
is not observed downstream from the sharp edge ramp. 
Furthermore, 
the shear layer growth is increased
as compared to the baseline case.
The GPC-controlled shear layer 
displays a mean streamwise velocity similarity state beyond $x/L_{\mathrm{Sep}} = 0.25$. 
The mean vortex region downstream of the sharp edge ramp growths
monotonically for GPC-based closed-loop control
--- corroborating the rapid development of the shear layer.
As expected for high-frequency forcing, GPC 
yields a finer distribution of vortex region population
as compared to baseline or best open-loop control.
Summarizing, GPC yields similar actuation benefits as best open-loop control
but at lower momentum coefficient.
This improvement is based on distinctly different high-frequency flow structures,
not on the lock-on of the periodic actuation response.

\section*{acknowledgements}
This work was supported by French National Research Agency (ANR) 
via the SepaCoDe Project (ANR-11-BS09-018) 
and the TUCOROM Chair of Excellence (ANR-10-CEXC-0015).
%\end{acknowledgements}

\bibliographystyle{spbasic}      % basic style, author-year citations
\bibliography{#main.bib#}

\begin{thebibliography}{39}
\providecommand{\natexlab}[1]{#1}
\providecommand{\url}[1]{{#1}}
\providecommand{\urlprefix}{URL }
\expandafter\ifx\csname urlstyle\endcsname\relax
  \providecommand{\doi}[1]{DOI~\discretionary{}{}{}#1}\else
  \providecommand{\doi}{DOI~\discretionary{}{}{}\begingroup
  \urlstyle{rm}\Url}\fi
\providecommand{\eprint}[2][]{\url{#2}}

\bibitem[{Amitay and Glezer(2002)}]{Amitay2002AIAA}
Amitay M, Glezer A (2002) Role of actuation frequency in controlled flow
  reattachment over a stalled airfoil. AIAA J 40(2):209--216

\bibitem[{Benard et~al(2010)Benard, Moreau, Griffin, and
  Cattafesta~III}]{Benard2010EiF}
Benard N, Moreau E, Griffin J, Cattafesta~III LN (2010) Slope seeking for
  autonomous lift improvement by plasma surface discharge. Exp Fluids
  48(5):791--808

\bibitem[{Benedict and Gould(1996)}]{Benedict1996EiF}
Benedict LH, Gould RD (1996) Towards better uncertainty estimates for
  turbulence statistics. Exp Fluids 22(2):129--136

\bibitem[{Browand and Troutt(1985)}]{Browand1985}
Browand FK, Troutt TR (1985) The turbulent mixing layer: geometry of large
  vortices. J Fluid Mech 158:489--509

\bibitem[{Brunton and Noack(2015)}]{Brunton2015amr}
Brunton SL, Noack BR (2015) Closed-loop turbulence control: Progress and
  challenges. Appl\ Mech\ Rev (in print)

\bibitem[{Castro and Haque(1987)}]{Castro1987}
Castro IP, Haque A (1987) The structure of a turbulent shear layer bounding a
  separation region. J Fluid Mech 179:439--468

\bibitem[{Cherry et~al(1984)Cherry, Hillier, and Latour}]{Cherry1984JFM}
Cherry NJ, Hillier R, Latour MEMP (1984) Unsteady measurements in a separated
  and reattaching flow. J Fluid Mech 144:13--46

\bibitem[{Cordier et~al(2013)Cordier, Noack, Tissot, Lehnasch, Delville,
  Balajewicz, Daviller, and
  Niven}]{Cordier_Noack_Tissot_Lehnasch_Delville_Balajewicz_Daviller_Niven_2013_EIF}
Cordier L, Noack BR, Tissot G, Lehnasch G, Delville J, Balajewicz M, Daviller
  G, Niven RK (2013) Identification strategy for model-based control. Exp in
  Fluids 54:1580

\bibitem[{Cuvier et~al(2011)Cuvier, Braud, Foucault, and
  Stanislas}]{Cuvier2011TSFP7b}
Cuvier C, Braud C, Foucault JM, Stanislas M (2011) Flow control over a ramp
  using active vortex generators. In: Seventh International Symposium on
  Turbulence and Shear Flow Phenomena, Ottawa, Canada

\bibitem[{Dandois et~al(2007)Dandois, Garnier, and Sagaut}]{Dandois2007JFM}
Dandois J, Garnier E, Sagaut P (2007) Numerical simulation of active separation
  control by a synthetic jet. J Fluid Mech 574(1):25--58

\bibitem[{Debien et~al(2014)Debien, Aubrun, Mazellier, and
  Kourta}]{Debien2014CRAS}
Debien A, Aubrun S, Mazellier N, Kourta A (2014) Salient and smooth edge ramps
  inducing turbulent boundary layer separation: Flow characterization for
  control perspective. C R M{\'e}canique

\bibitem[{Debien et~al(2015)Debien, Aubrun, Mazellier, and
  Kourta}]{Debien2015TSFP9}
Debien A, Aubrun S, Mazellier N, Kourta A (2015) Active separation control
  process over a sharp edge ramp. In: Ninth International Symposium on
  Turbulence and Shear Flow Phenomena, Melbourne, Australia

\bibitem[{Duriez et~al(2014)Duriez, Parezanovi\'c, Laurentie, Fourment,
  Delville, Bonnet, Cordier, Noack, Segond, Abel, Gautier, Aider, Raibaudo,
  Cuvier, Stanislas, and Brunton}]{Duriez2014aiaa}
Duriez T, Parezanovi\'c V, Laurentie JC, Fourment C, Delville J, Bonnet JP,
  Cordier L, Noack BR, Segond M, Abel MW, Gautier N, Aider JL, Raibaudo C,
  Cuvier C, Stanislas M, Brunton S (2014) {Closed-loop control of experimental
  shear layers using machine learning (Invited)}. In: 7th AIAA Flow Control
  Conference, Atlanta, Georgia, USA, pp 1--16

\bibitem[{Gautier and Aider(2013)}]{Gautier2013}
Gautier N, Aider JL (2013) Control of the separated flow downstream of a
  backward-facing step using visual feedback. Proc R Soc A 469(2160):20130,404

\bibitem[{Gautier et~al(2015)Gautier, Aider, Duriez, Noack, Segond, and
  Abel}]{Gautier2015jfm}
Gautier N, Aider JL, Duriez T, Noack BR, Segond M, Abel M (2015) Closed-loop
  separation control using machine learning. J Fluid Mech 770:442--457

\bibitem[{Gerhard et~al(2003)Gerhard, Pastoor, King, Noack, Dillmann,
  Morzy\'nski, and Tadmor}]{Gerhard2003aiaa}
Gerhard J, Pastoor M, King R, Noack BR, Dillmann A, Morzy\'nski M, Tadmor G
  (2003) Model-based control of vortex shedding using low-dimensional
  {G}alerkin models. In: 33rd AIAA Fluids Conference and Exhibit, Orlando,
  Florida, USA, paper 2003-4262

\bibitem[{Godard and Stanislas(2006{\natexlab{a}})}]{Godard2006ASTP1}
Godard G, Stanislas M (2006{\natexlab{a}}) {Control of a decelerating boundary
  layer. Part 1: Optimization of passive vortex generators}. Aerosp Sci Technol
  10(3):181--191

\bibitem[{Godard and Stanislas(2006{\natexlab{b}})}]{Godard2006ASTP3}
Godard G, Stanislas M (2006{\natexlab{b}}) {Control of a decelerating boundary
  layer. Part 3: Optimization of round jets vortex generators}. Aerosp Sci
  Technol 10(6):455--464

\bibitem[{Greenblatt and Wygnanski(2000)}]{Greenblatt2000PAS}
Greenblatt D, Wygnanski IJ (2000) The control of flow separation by periodic
  excitation. Prog Aerosp Sci 36(7):487--545

\bibitem[{Hasan(1992)}]{Hasan1992JFM}
Hasan MAZ (1992) The flow over a backward-facing step under controlled
  perturbation: laminar separation. J Fluid Mech 238:73--96

\bibitem[{Jovic(1996)}]{Jovic1996}
Jovic S (1996) {An experimental study of a separated/reattached flow behind a
  backward-facing step. $Re_H = 37, 000$}. Tech. Mem. 110384, NASA

\bibitem[{King(2007)}]{King2007book}
King R (ed) (2007) {Active Flow Control I}, no.~95 in Notes on Numerical Fluid
  Mechanics and Interdisciplinary Design, Springer-Verlag, Berlin

\bibitem[{King(2010)}]{King2010book}
King R (ed) (2010) {Active Flow Control II}, no. 108 in Notes on Numerical
  Fluid Mechanics and Interdisciplinary Design, Springer-Verlag, Berlin

\bibitem[{Koza(1992)}]{Koza1992book}
Koza JR (1992) {Genetic Programming: On the Programming of Computers by Means
  of Natural Selection}. The MIT Press, Boston

\bibitem[{Koza et~al(1999)Koza, Bennett~III, and Stiffelman}]{Koza1999book}
Koza JR, Bennett~III FH, Stiffelman O (1999) {Genetic Programming as a
  Darwinian Invention Machine}, Lecture Notes in Computer Science, vol 1598.
  Springer

\bibitem[{Lin(2002)}]{Lin2002PAS}
Lin JC (2002) Review of research on low-profile vortex generators to control
  boundary-layer separation. Prog Aerosp Sci 38(4):389--420

\bibitem[{Mabey(1972)}]{Mabey1972JoA}
Mabey DG (1972) Analysis and correlation of data on pressure fluctuations in
  separated flow. J Aircraft 9(9):642--645

\bibitem[{Mittal et~al(2005)Mittal, Kotapati, and Cattafesta~III}]{Mittal2005}
Mittal R, Kotapati R, Cattafesta~III L (2005) {Numerical Study of Resonant
  Interactions and Flow Control in a Canonical Seperated Flow}. In: AIAA 43rd
  Aerospace Sciences Meeting and Exhibit, Reno, NV, USA

\bibitem[{Murphy(2012)}]{Murphy_2012}
Murphy KP (2012) Machine Learning: A Probabilistic Perspective. MIT Press,
  Cambridge

\bibitem[{Parezanovic et~al(2015)Parezanovic, Laurentie, Fourment, Delville,
  Bonnet, Spohn, Duriez, Cordier, Noack, Abel, Segond, Shaqarin, and
  Brunton}]{Parezanovic2015ftac}
Parezanovic V, Laurentie JC, Fourment C, Delville J, Bonnet JP, Spohn A, Duriez
  T, Cordier L, Noack BR, Abel M, Segond M, Shaqarin T, Brunton S (2015) Mixing
  layer manipulation experiment -- from open-loop forcing to closed-loop
  machine learning control. Flow Turbul Combust 94:155--173

\bibitem[{Pastoor et~al(2008)Pastoor, Henning, Noack, King, and
  Tadmor}]{Pastoor2008JFM}
Pastoor M, Henning L, Noack BR, King R, Tadmor G (2008) Feedback shear layer
  control for bluff body drag reduction. J Fluid Mech 608:161--196

\bibitem[{Petz et~al(2009)Petz, Kasten, Prohaska, and Hege}]{Petz2009}
Petz C, Kasten J, Prohaska S, Hege HC (2009) {Hierarchical Vortex Regions in
  Swirling Flow}. Comput Graph Forum 28(3):863--870

\bibitem[{Rowley and Williams(2006)}]{Rowley2006arfm}
Rowley CW, Williams DR (2006) Dynamics and control of high-{R}eynolds number
  flows over open cavities. Annu\ Rev\ Fluid Mech 38:251--276

\bibitem[{Seifert and Pack(2003)}]{Seifert2003JoA}
Seifert A, Pack LG (2003) {Effects of sweep on active separation control at
  high Reynolds numbers}. J Aircraft 40(1):120--126

\bibitem[{Shaqarin et~al(2011)Shaqarin, Braud, Coudert, and
  Stanislas}]{Shaqarin2011TSFP7}
Shaqarin T, Braud C, Coudert S, Stanislas M (2011) Open and closed-loop
  experiments to reattach a thick turbulent boundary layer. In: Seventh
  International Symposium on Turbulence and Shear Flow Phenomena, Ottawa,
  Canada

\bibitem[{Shaqarin et~al(2013)Shaqarin, Braud, Coudert, and
  Stanislas}]{Shaqarin2013EiF}
Shaqarin T, Braud C, Coudert S, Stanislas M (2013) Open and closed-loop
  experiments to identify the separated flow dynamics of a thick turbulent
  boundary layer. Exp Fluids 54(2):1--22

\bibitem[{Song et~al(2000)Song, DeGraaff, and Eaton}]{Song2000IJHFF}
Song S, DeGraaff DB, Eaton JK (2000) Experimental study of a separating,
  reattaching, and redeveloping flow over a smoothly contoured ramp. Int J Heat
  Fluid Fl 21(5):512--519

\bibitem[{Tennekes and Lumley(1972)}]{Tennekes1972MIT}
Tennekes H, Lumley J (1972) {A first course in Turbulence}. MIT Press

\bibitem[{Zaman and Hussain(1981)}]{Zaman1981JFM}
Zaman KBMQ, Hussain AKMF (1981) Turbulence suppression in free shear flows by
  controlled excitation. J Fluid Mech 103:133--159

\end{thebibliography}
%./biblio_article_book_util,./biblio_conference_util}
\end{document}